\documentclass[twocolumn,superscriptaddress,amsmath, amssymb, amsfonts,preprintnumbers,aps,prd,longbibliography,
nofootinbib]
{revtex4-1}

\usepackage{graphicx}% Include figure files
\usepackage{dcolumn}% Align table columns on decimal point
\usepackage{bm}% bold math
 \usepackage{amstext}
 \usepackage{amssymb}
 \usepackage{amsmath}
 \usepackage{graphicx}
 \usepackage{tikz}
\usepackage{tikz-feynman}
 \usepackage{color}
 \usepackage{bbold}
 \usepackage{delimset} % for brackets
\usepackage[caption=false,justification=justified]{subfig}
\usepackage[colorlinks=true]{hyperref}
\usepackage{orcidlink}
\usepackage{float}
\usepackage{soul}

\def\deltabar{{\mathchar '26\mkern -10mu\delta}}

\newcommand{\zenodo}{\href{https://doi.org/10.5281/zenodo.15740904}{\tt zenodo.org} \cite{zenodo} }

\DeclareFontFamily{OT1}{pzc}{}
\DeclareFontShape{OT1}{pzc}{m}{it}{<-> s * [1.350] pzcmi7t}{}
\DeclareMathAlphabet{\mathpzc}{OT1}{pzc}{m}{it}

\def\dd{\delta\!\!\!{}^-\!}

\def\d{\mathrm{d}}
\def\eps{\epsilon}

\def\braket#1{\langle #1 \rangle}

\def\nn{\nonumber}

\newcommand{\vev}[1]{\langle #1\rangle}

\newcommand{\be}{\begin{equation}}
\newcommand{\ee}{\end{equation}}
\newcommand{\ba}{\begin{align}}
\newcommand{\ea}{\end{align}}
\ifx\genfrac\sdflkaj\else\fi
\newcommand{\sfrac}[2]{{\textstyle\frac{#1}{#2}}}

\newcommand{\Z}{\mathcal{Z}}

\newcommand\scalemath[2]{\scalebox{#1}{\mbox{\ensuremath{\displaystyle #2}}}}

\def\eps{\epsilon}

\def\d{\mathrm{d}}

\usepackage{tikz}
\usetikzlibrary{decorations.pathmorphing}
\usetikzlibrary{decorations.markings}
\usetikzlibrary{positioning, shapes, snakes, arrows}
\usetikzlibrary{patterns}

\tikzset{
    wl/.style={line width=1pt},
    graviton/.style={line width=.8pt, -latex,decorate, decoration={snake, segment length=4pt,amplitude=1.8pt, pre length=.15cm, post length=.25cm}},
    worldlineStatic/.style={dotted, line width=1pt},
	worldline/.style={gray, line width=1pt},
	worldlineBold/.style={black, line width=.6pt},
	zUndirected/.style={line width=1pt},
	zParticle/.style={line width=1pt,postaction={decorate},decoration={markings,mark=at position .6 with {\arrow[#1]{latex}}}},
	zParticleF/.style={line width=1pt,postaction={decorate}},
	cscalar/.style={line width=1pt,postaction={decorate},decoration={markings,mark=at position .6 with {\arrow[#1]{latex}}}},
	cscalar2/.style={line width=1pt,postaction={decorate},decoration={markings,mark=at position .8 with {\arrow[#1]{latex}}}},
	photon/.style={line width =.8pt, decorate, decoration={snake, segment length=4pt, amplitude=1.8pt,  pre length=.1cm, post length=.1cm}},
	photonRed/.style={red, line width =.8pt, decorate, decoration={snake, segment length=4pt, amplitude=1.8pt,  pre length=.1cm, post length=.1cm}},
	cross/.style={cross out, line width =.8pt, draw=black, minimum size=2*(#1-\pgflinewidth), inner sep=0pt, outer sep=0pt},
cross/.default={4pt}
}

\def\mn{{\mu\nu}}

\def\i\math

\def\bH{\hat{b}}

\def\dd{\delta\!\!\!{}^-\!}

\def\d{\mathrm{d}}
\def\eps{\epsilon}

\def\braket#1{\langle #1 \rangle}

\renewcommand{\i}{\ensuremath{\mathrm{i}}}
\renewcommand{\d}{\ensuremath{\mathrm{d}}}

\def\cO{\mathcal{O}}

\def\cZ{\mathcal{Z}}

\def\eps{\epsilon}

\def\d{\mathrm{d}}

\def\nn{\nonumber}

\def\eqn#1{eq.~\eqref{#1}}

\def\Rcite#1{Ref.~\cite{#1}}

\DeclareMathOperator{\arccosh}{arccosh}

\begin{document}

\allowdisplaybreaks

\preprint{HU-EP-25/20-RTG, QMUL-PH-25-10}

\title{Radiated Angular Momentum from Spinning Black Hole Scattering Trajectories}

\author{Gustav Mogull\,\orcidlink{0000-0003-3070-5717}}
\email{g.mogull@qmul.ac.uk}
\affiliation{%
Institut f\"ur Physik, Humboldt-Universit\"at zu Berlin,
10099 Berlin, Germany
}
\affiliation{Max Planck Institut f\"ur Gravitationsphysik (Albert Einstein Institut), 14476 Potsdam, Germany}
\affiliation{Centre for Theoretical Physics, Department of Physics and Astronomy, Queen Mary University of London,  London E1~4NS, United Kingdom}

\author{Jan Plefka\,\orcidlink{0000-0003-2883-7825}} 
\email{jan.plefka@hu-berlin.de}
\affiliation{%
Institut f\"ur Physik, Humboldt-Universit\"at zu Berlin,
10099 Berlin, Germany
}

\author{Kathrin Stoldt\,\orcidlink{0009-0001-4858-5218}}
\email{kathrin.stoldt.1@hu-berlin.de}
\affiliation{%
Institut f\"ur Physik, Humboldt-Universit\"at zu Berlin,
10099 Berlin, Germany
}

\begin{abstract}
Using the worldline quantum field theory approach we derive solutions to the equations
of motion for spinning massive bodies up to quadratic order in spins.
At leading post-Minkowskian (PM) order these trajectories are obtained in the time domain,
and at sub-leading order in the frequency domain.
Our approach incorporates diagrammatic techniques and modern Feynman integration technologies,
and includes a
%new
 family of loop integrals different to those seen in asymptotic PM calculations. Our results provide a new mechanism for computing
the radiated angular momentum involved in gravitational scattering,
which we reproduce at 2PM order up to linear spins.
We have established a framework for computing higher-order effects to further extend the
high-precision frontier in analytical gravitational wave physics,
and push predictions for the radiated angular momentum to higher perturbative orders.
% We compute radiated angular momentum in spinning black hole scattering using worldline quantum field theory. While recent advances have yielded high-precision results for various scattering observables, the direct calculation of 
% radiated angular momentum has remained elusive. Here, we derive explicit solutions to the equations of motion for spinning massive
% bodies at 2PM accuracy in the frequency domain. Using these trajectories, we calculate the radiated angular momentum at order $G^{2}$
% for the spinning case. Our approach incorporates diagrammatic techniques and modern Feynman integration technologies.
% It establishes a framework for computing higher-order effects to further extend the high-precision frontier in analytical gravitational
% wave physics.
\end{abstract}

\maketitle

\section{Introduction}

The gravitational two-body problem has held a foundational position in physics 
since the days of Newton.
In its general relativistic version it not only represents a fascinating intellectual challenge ---
given its conceptual simplicity paired with its technical complexity ---
but has become increasingly important due to the advent of gravitational wave astronomy \cite{LIGOScientific:2016aoc}.
Today, gravitational wave detectors routinely observe gravitational waves emitted from binary black hole and neutron 
star systems \cite{KAGRA:2021vkt} and future third generation observatories will profoundly advance this in the coming 
decade \cite{LISA:2017pwj,Punturo:2010zz,Ballmer:2022uxx,Abac:2025saz}, calling 
for high-precision theoretical predictions from Einstein gravity to enable data analysis and interpretation.

Within perturbative frameworks to address the scattering variant of the relativistic two-body 
problem an expansion in Newton’s constant $G$ --- the post-Minkowskian (PM) expansion ---
is most natural~\cite{Kovacs:1978eu,Westpfahl:1979gu,Bel:1981be,Damour:2017zjx,Hopper:2022rwo}. 
Here, dramatic progress in the past years was made by translating this
classical physics problem to a scattering problem in perturbative
quantum field theory (QFT), thereby unleashing state-of-the-art techniques developed for
collider physics~\cite{Kalin:2020mvi,Kalin:2020fhe,Mogull:2020sak,Dlapa:2021npj,Dlapa:2021vgp,Riva:2021vnj,Dlapa:2022lmu,Liu:2021zxr,Dlapa:2023hsl,Jakobsen:2021smu,Jakobsen:2021zvh,Jakobsen:2022psy,Luna:2017dtq,Kosower:2018adc,Bjerrum-Bohr:2018xdl,Bern:2019nnu,Bern:2019crd,Cristofoli:2020uzm,Bjerrum-Bohr:2021wwt,Cheung:2020gyp,Bjerrum-Bohr:2021din,DiVecchia:2020ymx,DiVecchia:2021bdo,DiVecchia:2021ndb,DiVecchia:2022piu,DiVecchia:2023frv,Heissenberg:2022tsn,Damour:2020tta,Herrmann:2021tct,Damgaard:2019lfh,Damgaard:2019lfh,Damgaard:2021ipf,Damgaard:2023vnx,Aoude:2020onz,AccettulliHuber:2020dal,Brandhuber:2021eyq,Bern:2021dqo,Bern:2021yeh,Damgaard:2023ttc}. 
In doing so we today have control up to the fifth order in PM
perturbation theory (5PM) for the scattering observables of the scattering angle and 
radiated four-momentum~\cite{Driesse:2024xad,Driesse:2024feo} (see Refs.~\cite{Bern:2023ccb,Bern:2024adl} for related work in electrodynamics and supergravity), as well as up to the next-to-leading order for the far-field
scattering 
waveform~\cite{Jakobsen:2021smu,Mougiakakos:2021ckm,Jakobsen:2021lvp,Cristofoli:2021vyo,Brandhuber:2023hhy,Brandhuber:2023hhy,Brandhuber:2023hhl,DeAngelis:2023lvf,Herderschee:2023fxh,Caron-Huot:2023vxl,Bohnenblust:2023qmy,Bini:2024rsy,Brunello:2024ibk,Elkhidir:2023dco,Georgoudis:2023lgf,Bohnenblust:2025gir}. Importantly, spin degrees of freedom may also be included in these effective field
theory descriptions ~\cite{Vines:2017hyw,Bern:2022kto,Aoude:2023vdk,Bern:2023ity,FebresCordero:2022jts,Bern:2023ity,Alaverdian:2024spu,Alaverdian:2025jtw,Bohnenblust:2024hkw,Jakobsen:2021lvp,Jakobsen:2021zvh,Haddad:2024ebn,Bonocore:2025stf} and have been
similarly advanced to the 5PM order (in a physical PM counting) \cite{Jakobsen:2022fcj,Jakobsen:2022zsx,Jakobsen:2023ndj,Jakobsen:2023hig}.

These QFT-inspired approaches are designed to compute asymptotic quantities:
 the change in momentum or spins of bodies involved in the scattering process,
scattering angles, and the far-field waveform that encodes radiated linear and angular momentum
carried by the gravitational wave pulse.
Of these, the \emph{radiated angular momentum} is one of the more intricate to obtain~\cite{Manohar:2022dea}:
current methods include extracting it from the far-field waveform~\cite{Damour:2020tta,Jakobsen:2021smu,Jakobsen:2021lvp},
linear response relations~\cite{Bini:2012ji,Damour:2020tta,Bini:2021gat,Jakobsen:2022zsx,Jakobsen:2023hig}
or the eikonal~\cite{Heissenberg:2023uvo,Heissenberg:2024umh,Heissenberg:2025ocy}.
More recently, the use of Dirac brackets on an exponential representation of the $S$-matrix
has also been explored~\cite{Alessio:2025flu}.
So far, the explicit form of the full time-dependent \emph{trajectories} has also not been obtained with QFT methods.
Current state of-the-art is 2PM order,
derived by integrating the equations of motion directly~\cite{Westpfahl:1969ea,Westpfahl:1979gu,Westpfahl:1980mk,Bel:1981be,Westpfahl:1985tsl,Bini:2018ywr,Bini:2022wrq,Bini:2024hme}.

Here, we employ the worldline quantum field theory (WQFT)
approach~\cite{Mogull:2020sak,Jakobsen:2021zvh,Jakobsen:2022psy,Jakobsen:2023oow,Haddad:2024ebn}
to directly generate solutions to the equations of motion of the gravitational two-body problem through diagrammatic techniques, thereby also including spinning degrees of freedom for the first time.
At leading 1PM order we provide explicit results up to quadratic order in spins in the time domain,
and at sub-leading 2PM order we provide linear-in-spin results in Fourier space. 
% frequency domain.
Using these new results, we are able to reproduce the radiated angular momentum at
order $G^{2}$.
Under the intrinsic choice of BMS frame~\cite{Veneziano:2022zwh},
the radiated angular momentum starts at $G^{2}$ while
the radiated momentum is delayed to $G^{3}$~\cite{Damour:1981bh,Damour:2020tta}.

Our results are achieved by incorporating integration-by-parts,
differential equations and the method of regions techniques from the modern technical toolbox of collider physics.
Interestingly, we encounter a family of one-loop Feynman integrals that is a subset of Feynman integrals contributing to the gravitational wave form at $\mathcal{O}(G^3)$ \cite{Caron-Huot:2023vxl, Bohnenblust:2023qmy, Brandhuber:2023hhy, Herderschee:2023fxh}
 %have not been studied
%in the QFT literature on gravitational scattering
.
Our work builds on classical works for the leading-order dynamics~\cite{Westpfahl:1969ea,Westpfahl:1979gu,Westpfahl:1980mk,Bel:1981be,Westpfahl:1985tsl,Bini:2018ywr} and in particular
recent 2PM works \cite{Bini:2022wrq,Bini:2024hme}
that established and solved the equations of motion in the non-spinning case.

\section{Classical black hole scattering trajectories}

Our starting point is an effective description of spinning black holes
as point particles, described by a one-dimensional worldline action modelling the spins via bosonic oscillators~\cite{Haddad:2024ebn}.
This approach is superior to the previous $\mathcal{N}=2$ supersymmetric formalism \cite{Jakobsen:2021zvh}, as it is not restricted to quadratic spin order and is more practicable for not needing to deal with anti-commuting objects.
Up to terms quadratic in spin, the $i$'th black hole action is ($i=1,2$)
\begin{align}\label{SWQFT}
    &\frac{S^{(i)}}{m_i}\\
    &=-\!\int\!\d\tau_i \bigg[\sfrac{1}{2}g_{\mu\nu}\dot{x}_i^\mu\dot{x}_i^\nu\!+\!\i \bar{\alpha}_{i,\mu}\frac{{\rm D}\alpha^{\mu}_{i}}{\d\tau_{i}}
    \!+\!\sfrac12R_{\mu\nu\rho\sigma}\bar{\alpha}_i^\mu\alpha_i^\nu\bar{\alpha}_i^\rho\alpha_i^\sigma\bigg].\nn
\end{align}
The worldline trajectory $x_{i}^{\mu}(\tau_i)$ is augmented by a complex vector 
$\alpha_{i}^{\mu}(\tau_i)$, which is a bosonic oscillator mode.
Here $\frac{{\rm D}\alpha^{\mu}_{i}}{\d\tau_{i}}=\dot{\alpha}_{i}^{\mu}+\Gamma^{\mu}{}_{\nu\rho}\dot{x}_i^\nu   \alpha^\rho_i$ is the covariant worldline derivative employing 
the Levi-Civita connection.
These oscillators encode the spin tensor of the compact object~\cite{Haddad:2024ebn}:
\begin{align}
\label{spintensor}
    S_{i}^{\mu\nu}(\tau_{i})=-2m_i\i\, \bar\alpha_{i}^{[\mu}(\tau_i)\alpha_{i}^{\nu]}(\tau_i)\,,
\end{align}
anti-symmetrizing with unit weight.
The time-dependent spin vector is correspondingly defined by
\begin{align}\label{SVec}
     S_i^\mu (\tau_{i})&=\sfrac12{\eps^\mu}_{\nu\rho\sigma}S_i^{\nu\rho}(\tau_i)\dot{x}_i^\sigma(\tau_i)\,.
\end{align}
We consider scattering only of Kerr black holes and not spinning neutron stars,
which would necessitate an additional quadratic-in-spin term~\cite{Haddad:2024ebn}.
In \eqn{SWQFT} we have adopted a generalized proper time gauge in which
\begin{equation}\label{onshellcond}
g_{\mu\nu}\dot{x}_i^\mu\dot{x}_i^\nu=1-\frac1{4m_i^2}R_{\mu\nu\rho\sigma}S_i^{\mu\nu}S_i^{\rho\sigma}\, .
\end{equation}
Inclusion of the quadratic-in-spin term implies that
this is \emph{not} a proper-time gauge.

The complete two-body system is described by an action comprising the two black holes,
coupled to bulk gravity and including a de Donder gauge-fixing term for the graviton:
% \begin{equation}\label{saction}
%     S=- \frac{1}{16\pi G} \int\!\d^4x \sqrt{-g}\, R[g] +S_{\text{gf}}+\sum_{i=1}^2 S^{(i)} .
% \end{equation}
\begin{equation}\label{saction}
    S=-\frac2{\kappa^2}\int\!\d^Dx\sqrt{-g} \, R[g]+ S_{\text{g.f.}}
    %(\partial_\nu h^{\mu\nu}-\sfrac12\partial^\mu{h^\nu}_\nu)^2\big)
    +\sum_{i=1}^2 S^{(i)}\,,
\end{equation}
where $\kappa=\sqrt{32\pi G}$.
Anticipating the need for dimensional regularisation,
we work in $D=4-2\eps$ dimensions.
As we are interested in scattering solutions to the classical equations of motion,
we perform the background-field expansions:
\begin{subequations}\label{bgexpansions}
\begin{align}\label{trajed}
x_i^{\mu}(\tau_i)&= b_i^\mu + v_i^\mu\tau_i+ z_i^\mu(\tau_i), \\
\alpha_{i}^{\mu}(\tau_i)& = \alpha_{-\infty,i}^{ \mu}  + \alpha_{i}^{\prime\, \mu}(\tau_i)\, , \\
S_{i}^{\mu}(\tau_i)& = m_i(a_i^\mu  + a_{i}^{\prime\, \mu}(\tau_i))\, , \\
g_{\mu\nu}(x)&= \eta_{\mu\nu} + \kappa\, h_{\mu\nu}(x) \, .
\end{align}
\end{subequations}
Moreover $z_i^\mu(\tau_i)$, $\alpha_{i}^{\prime\, \mu}(\tau_i)$, $a_{i}^{\prime\, \mu}(\tau_i)$
and $\kappa\, h_{\mu\nu}(x)$ are expanded in $G$ according to
\begin{subequations}\label{backgroundExp}
\begin{align}
z_i^\mu(\tau_i)&=G z_i^{(1)\mu}(\tau_i)+G^2 z_i^{(2)\mu}(\tau_i)+\mathcal{O}(G^3)\, ,\\
\alpha_{i}^{\prime\,\mu}(\tau_i)&=G \alpha_i^{(1)\mu}(\tau_i)+G^2 \alpha_i^{(2)\mu}(\tau_i)+\mathcal{O}(G^3)\, ,\\
a_{i}^{\prime\,\mu}(\tau_i)&=G a_i^{(1)\mu}(\tau_i)+G^2 a_i^{(2)\mu}(\tau_i)+\mathcal{O}(G^3)\, ,\\
\kappa\, h_{\mu\nu}(x)&=G h_{\mu\nu}^{(1)}(x)+G^2 h_{\mu\nu}^{(2)}(x)+\mathcal{O}(G^3)\, .
\end{align}
\end{subequations}
In terms of the background variables, all defined at past infinity,
we have the relative impact parameter $b^\mu=b_2^\mu-b_1^\mu$,
incoming momenta $p_i^\mu=m_iv_i^\mu$, Pauli-Lubanski spin vectors $a_i^\mu$,
and the initial spin tensor
$S_{-\infty,i}^{\mu\nu}=-2m_i\i\bar\alpha_{-\infty,i}^{[\mu}\alpha_{-\infty,i}^{\nu]}$.
The incoming velocities satisfy $v_1^2=v_2^2=1$,
and $\gamma=v_1\cdot v_2=1/\sqrt{1-v^2}$ is the dimensionless boost factor.
There is also the unit-normalized angular momentum vector
\begin{align}
   \hat{l}^\mu=-{\eps^\mu}_{\nu\rho\sigma}\frac{b^\nu v_1^\rho v_2^\sigma}{|b|\sqrt{\gamma^2-1}}\,,
\end{align}
and the dual velocities,
\begin{align}
w^\mu_1&:=\frac{\gamma v_2^\mu-v_1^\mu}{\gamma^2-1}\,, & w_2^\mu&:=\frac{\gamma v_1^\mu-v_2^\mu}{\gamma^2-1}\,.
\end{align}
These definitions ensure that $v_i\cdot w_j=\delta_{ij}$.

Using retarded worldline propagators fixes the background parameters such that the black holes are incoming from past infinity. This will lead to a logarithmic divergence in the trajectory, which we subtract off later on.
In order to efficiently generate perturbative solutions to the classical equations of motion,
we use the WQFT approach~\cite{Mogull:2020sak,Jakobsen:2022psy},
exploiting the fact that the tree-level one-point functions of $\vev{z_i^\mu}$ and $\vev{h_{\mu\nu}}$
solve the classical equations of motions.
We will subsequently drop the expectation value notation for convenience.
In WQFT, perturbative post-Minkowskian (PM) classical solutions are represented as sums over tree-level Feynman diagrams.
The graviton propagator (in de Donder gauge) is denoted by a wiggly line:
\begin{equation}
\raisebox{0.3cm}{
\begin{tikzpicture}[baseline={(current bounding box.center)}]
\begin{feynman}
\vertex at ($(0,0)$) (a);
\vertex at ($(1.5,0)$) (b);
\diagram*{
a--[photon,thick, momentum=$k$] (b),
};
\draw [fill] (a) circle (.06) node [above] {$\mu$} node [below] {$\nu$};
\draw [fill] (b) circle (.06) node [above] {$\rho$} node [below] {$\sigma$};
\end{feynman}
\end{tikzpicture}}=\i\frac{\eta_{\mu(\rho}\eta_{\sigma)\nu}-\sfrac{1}{D-2}\eta_{\mu\nu}\eta_{\rho\sigma}}{(k^{0}+\i 0^{+})^2-\mathbf{k}^2} \, .
\end{equation}
For the two worldline fields in energy space, $z_{i}^{\mu}(\omega)$ and 
$\alpha_{i}^{\prime \mu}(\omega)$, we collect these in a composite vector \cite{Haddad:2024ebn},
\begin{align}
 \Z^{\mu}_{I}(\omega)=\{z^{\mu}(\omega), \alpha^{\prime \mu}(\omega),
\bar\alpha^{\prime \mu}(\omega)\}
\, ,
\end{align}
with flavour index $I$.
The retarded propagator of this flavoured worldline field then reads 
\begin{align}\label{WLprops}
 \begin{tikzpicture}[baseline={(current bounding box.center)}]
    \coordinate (in) at (-0.6,0);
    \coordinate (out) at (1.4,0);
    \coordinate (x) at (-.2,0);
    \coordinate (y) at (1.0,0);
    \draw [zParticle] (x) -- (y) node [midway, below] {$\omega$};
    \draw [dotted] (in) -- (x);
    \draw [dotted] (y) -- (out);
    \draw [fill] (x) circle (.06) node [above] {$\mu$} node [below] {$I$};
    \draw [fill] (y) circle (.06) node [above] {$\nu$} node [below] {$J$};
  \end{tikzpicture}=%\langle \Z_{I}^{\mu}(-\omega) \Z^{\nu}_{J}(\omega) \rangle_{0}=
-\i\frac{\eta^{\mu\nu}}{m}
 \left (\begin{matrix}
  \frac1{(\omega+\i0)^{2}} & 0 & 0 \\
  0 & 0 &   -\frac1{\omega+\i0} \\
0 & \frac1{\omega+\i0} & 0
 \end{matrix}\right )_{IJ}.
\end{align}
As dictated by causality and the in-in formalism \textcolor{red}{\cite{Jakobsen:2022psy, Galley:2012hx, Kalin:2022hph}},
we exclusively use retarded worldline propagators. 
In this diagram, the arrow indicates causality flow and
the energy $\omega$ travels in the same direction.
The Feynman vertices emerging from Eq.~\eqref{saction} have been
collected in Refs.~\cite{Mogull:2020sak,Driesse:2024feo} for the non-spinning case and in Ref.~\cite{Haddad:2024ebn} for the spinning case.
% For the convenience of the reader we collected the needed vertices in the
%appendix. \jan{maybe drop}\gustav{Yes, I would just refer to our recent paper with Kays.}\kathrin{I agree}

Working in momentum (energy) space, the full
trajectory of the $i$'th black hole is given by
\be\label{ztauom}
   \Z_i^\mu(\tau_i) =
   \int_\omega\! e^{-\i \omega \tau_i}\braket{\Z^{\mu}_i(\omega)}\,,
\ee
where we introduced the shorthand notation $\int_\omega:=\int\!\frac{\d\omega}{2\pi}$.
%\red{$$
%z_i^\mu(\tau_i) = \int_{\omega}\, e^{-\i \omega \tau} \vev{z^{\mu}_{i}(\omega)}\,.
%$$}
The one-point function $\vev{\Z^{\mu}_1(\omega)}$
is given by a sum over diagrams with a single outgoing line,
carrying energy $\omega$:
\begin{align}\label{eq:1PtFns}
    \langle \Z_1^\mu(\omega)\rangle=\vcenter{\hbox{\begin{tikzpicture}[line cap=round,line join=round,x=1cm,y=1cm]
        \node (A1) at (-1.33,.0) {$\Z_1^\mu(\omega)$};
        \node (A2) at (-2.5,-.3) {$\omega\rightarrow$};
        \node (A3) at (-4.4,-0.7) {$q\, \uparrow$};
        \path [draw=black, worldlineStatic] (-4.7,0) -- (-3.83,0);
        \path [draw=black, zParticle] (-3.17,0) -- (-2.2,0);
        \path [draw=black, worldlineStatic] (-4.7,-1.5) -- (-3.83,-1.5);
        \path [draw=black, worldlineStatic] (-3.17,-1.5) -- (-2.2,-1.5);
        %\path [draw=black, zUndirected] (-4.6,0) -- (-3.4,0);
        \draw[pattern=north east lines] (-3.5,-0.75) ellipse (.55cm and .9cm);
    \end{tikzpicture}}}\,.
\end{align}
In schematic form, the full trajectory of the first black hole (BH) is given by (we write 
$\int_q:=\int\!\frac{\d^{D}q}{(2\pi)^{D}}$)
\begin{align}\label{trajFourier} 
   \Z_1^\mu(\tau_1) &= \int_{q}\int_{\omega}\, e^{i(q\cdot b-\omega
    \tau_1)}\deltabar(q\cdot v_1-\omega)\deltabar(q\cdot v_2) \Z_1^{\mu}(\omega)\nn\\
   &=\int_{q}\, e^{iq\cdot\tilde b(\tau_1)}\deltabar(q\cdot v_2) \Z_{1}^{\mu}(q)\,,
\end{align}
where $q^\mu$ is the total momentum exchanged between the two worldlines
and we have introduced the shifted impact parameter:
\begin{align}\label{shiftedB}
   \tilde{b}^\mu(\tau_1):=b^\mu-v_1^\mu\tau_1\,.
\end{align}
This extends the relative impact parameter $b^\mu=b_2^\mu-b_1^\mu$ along
the undeflected trajectory of the first black hole.

\begin{figure}[t!]
   \begin{center}
   \scalemath{0.78}{
   \begin{tikzpicture}
   \begin{feynman}
   \vertex[dot] at ($(0,0)$) (a){};
   \vertex[dot] at ($(0,2)$) (b){};
   \vertex at ($(-1,0)$) (c) {2};
   \vertex[dot] at ($(1,0)$) (d){};
   \vertex[dot] at ($(1,2)$) (e){};
   \vertex at ($(2,0)$) (f);
   \vertex at ($(-1,2)$) (g) {1};
   \vertex at ($(2,2)$) (h);
   \diagram*{
   a--[dotted] (c),
   a--[dotted] (d),
   d--[dotted] (f),
   a--[photon, thick, momentum=\(\ell\)] (b),
   b--[dotted] (g),
   b--[fermion, thick, edge label=\(\tilde{\omega}\)] (e),
   e--[photon, thick, rmomentum=\(q-\ell\)] (d),
   e--[fermion, thick, edge label=\(\omega\)] (h),};
   \end{feynman}
   \end{tikzpicture}
   \qquad
   \begin{tikzpicture}
   \begin{feynman}
   \vertex at ($(0,0)$) (a);
   \vertex at ($(1.5,0)$) (b);
   \vertex[dot] at ($(-0.5,0)$) (c){};
   \vertex[dot] at ($(0.5,0)$) (d){};
   \vertex[dot] at ($(0,1)$) (e){};
   \vertex[dot] at ($(0,2)$) (f){};
   \vertex at ($(-1.5,2)$) (g) {1};
   \vertex at ($(1.5,2)$) (h);
   \vertex at (-1.5,0) (i) {2};
   \diagram*{
   (a)--[dotted] (c),
   (a)--[dotted] (d),
   (d)--[dotted] (b),
   (c)--[photon,thick,momentum=\(\ell\)] (e),
   (a)--[dotted](i),
   (e)--[photon, thick, rmomentum=\(q-\ell\)] (d),
   (e)--[photon,thick, momentum=\(q\)] (f),
   (f)--[dotted] (g),
   (f)--[fermion, thick, edge label=\(\omega\),thick] (h),};
   \end{feynman}
   \end{tikzpicture}
   \qquad
   \begin{tikzpicture}
   \begin{feynman}
   \vertex at ($(0,0)$) (a) {2};
   \vertex[dot] at ($(0.75,0)$) (b){};
   \vertex[dot] at ($(2.25,0)$) (c){};
   \vertex at ($(3,0)$) (d);
   \vertex at ($(0,2)$) (e) {1};
   \vertex[dot] at ($(1.5,2)$) (f){};
   \vertex at ($(3,2)$) (g);
   \diagram*{
   (a)--[dotted] (b),
   (b)--[dotted] (c),
   (b)--[photon, thick, momentum=\(\ell\)] (f),
   (c)--[dotted] (d),
   (e)--[dotted] (f),
   (f)--[photon, thick, rmomentum=\(q-\ell\)] (c),
   (f)--[fermion, thick, edge label=\(\omega\)] (g),
   };
   \end{feynman}
   \end{tikzpicture}}
   \end{center}
   \begin{center}
      \scalemath{0.75}{
      \begin{tikzpicture}
      \begin{feynman}
      \vertex[dot] at ($(0,0)$) (a){};
      \vertex[dot] at ($(0,2)$) (b){};
      \vertex at ($(-1,0)$) (c) {2};
      \vertex[dot] at ($(1,0)$) (d){};
      \vertex[dot] at ($(1,2)$) (e){};
      \vertex at ($(2,0)$) (f);
      \vertex at ($(-1,2)$) (g) {1};
      \vertex at ($(2,2)$) (h);
      \diagram*{
      a--[dotted] (c),
      a--[fermion, thick, edge label=\(\tilde{\omega}\)] (d),
      d--[dotted] (f),
      a--[photon, thick, rmomentum=\(-\ell\)] (b),
      b--[dotted] (g),
      b--[dotted] (e),
      e--[photon, thick, rmomentum=\(q-\ell\)] (d),
      e--[fermion, thick, edge label=\(\omega\)] (h),};
      \end{feynman}
      \end{tikzpicture}
      \qquad
      \begin{tikzpicture}
      \begin{feynman}
      \vertex[dot] at ($(0,0)$) (a){};
      \vertex at ($(1.5,0)$) (b);
      \vertex at ($(-0.5,0)$) (c);
      \vertex at ($(0.5,0)$) (d);
      \vertex[dot] at ($(0,1)$) (e){};
      %\vertex at ($(0,2)$) (f);
      \vertex at ($(-1.5,2)$) (g) {1};
      \vertex at ($(1.5,2)$) (h);
      \vertex at (-1.5,0) (i) {2};
      \vertex[dot] at ($(-0.5,2)$) (j){};
      \vertex[dot] at ($(0.5,2)$) (k){};
      \diagram*{
      %(a)--[dotted] (c),
      (a)--[photon, thick, momentum=\(q\)] (e),
      (c)--[dotted] (d),
      (d)--[dotted] (b),
      (c)--[dotted](i),
      (e)--[photon, thick, rmomentum=\(-\ell\)] (j),
      (e)--[photon, thick, momentum'=\(q-\ell\)] (k),
      (k)--[dotted] (g),
      (k)--[fermion, thick, edge label=\(\omega\)] (h),};
      \end{feynman}
      \end{tikzpicture}
      \qquad
      \begin{tikzpicture}
      \begin{feynman}
      \vertex at ($(0,0)$) (a) {2};
      \vertex at ($(0,2)$) (b) {1};
      \vertex[dot] at ($(0.5,2)$) (c){};
      \vertex[dot] at ($(2,2)$) (d){};
      \vertex[dot] at ($(1.25,0)$) (e){};
      \vertex at ($(3,0)$) (f);
      \vertex at ($(3,2)$) (h);
      \diagram*{
      (a)--[dotted] (e),
      (e)--[dotted] (f),
      (e)--[photon, thick, rmomentum=\(-\ell\)] (c),
      (e)--[photon, thick, momentum'=\(q-\ell\)] (d),
      (c)--[dotted] (d),
      (c)--[dotted] (b),
      (d)--[fermion, thick, edge label=\(\omega\)] (h),};
      \end{feynman}
      \end{tikzpicture}}
      \end{center}
      
      \begin{center}
      \scalemath{0.75}{
      \begin{tikzpicture}
      \begin{feynman}
      \vertex[dot] at ($(0,0)$) (a){};
      \vertex[dot] at ($(0,2)$) (b){};
      \vertex at ($(-1,0)$) (c) {2};
      \vertex at ($(1,0)$) (d);
      \vertex[dot] at ($(1,2)$) (e){};
      \vertex at ($(2,0)$) (f);
      \vertex at ($(3,0)$) (i);
      \vertex at ($(-1,2)$) (g) {1};
      \vertex[dot] at ($(2,2)$) (h){};
      \vertex at ($(3,2)$) (j);
      \diagram*{
      (a)--[dotted] (c),
      (a)--[dotted] (d),
      (d)--[dotted] (f),
      (f)--[dotted] (i),
      (a)--[photon, thick, momentum=\(q\)] (b),
      (b)--[dotted] (g),
      (b)--[fermion, thick, edge label=\(\tilde{\omega}\)] (e),
      (e)--[dotted] (h),
      (e)--[photon, thick, half right, momentum'=\(q-\ell\)] (h),
      (h)--[fermion, thick, edge label=\(\omega\)] (j),};
      \end{feynman}
      \end{tikzpicture}
      \qquad
      \begin{tikzpicture}
      \begin{feynman}
      \vertex[dot] at ($(0,0)$) (a){};
      \vertex[dot] at ($(0,2)$) (b){};
      \vertex at ($(-1,0)$) (c) {2};
      \vertex at ($(1,0)$) (d);
      \vertex[dot] at ($(1,2)$) (e){};
      \vertex at ($(2,0)$) (f);
      \vertex at ($(-1,2)$) (g) {1};
      \vertex at ($(2,2)$) (h);
      \diagram*{
      (a)--[dotted] (c),
      (a)--[dotted] (d),
      (d)--[dotted] (f),
      %(f)--[dotted] (i),
      (a)--[photon, thick, momentum=\(q\)] (b),
      (b)--[dotted] (g),
      (b)--[photon, thick, half right, momentum'={[arrow shorten=0.25, xshift=4mm, rotate=9]\(q-\ell\)}] (e),
      (e)--[fermion, thick, edge label={$\omega$}] (h),};
      \end{feynman}
      \end{tikzpicture}}
      \end{center}
   \caption{Diagrams contributing to the sub-leading scattering trajectory $\Z_1^{(2)\mu}(\tau)$,
   including spin.
   The three diagrams in the first row are proportional to $m_2^2$,
   and contribute in the probe limit $m_1\ll m_2$;
   the five diagrams in the second and third rows are proportional to $m_1m_2$.
   \label{subleadingDiags}}
   \end{figure}
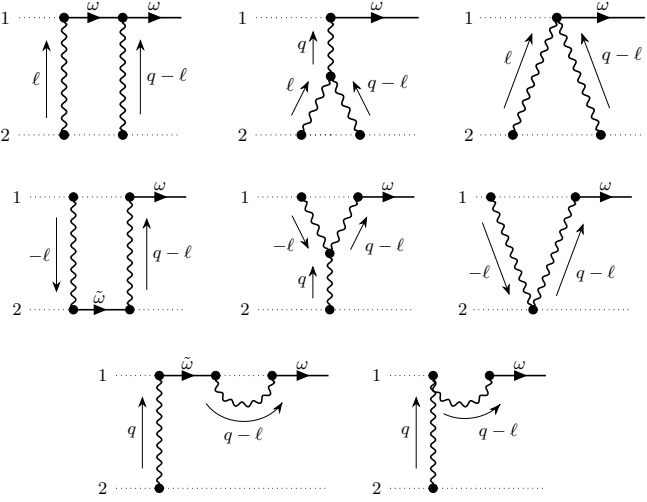

The diagrams involved in the trajectory contain those used to compute
the momentum impulse $\Delta p_1^\mu$ plus so-called ``mushroom’’ diagrams, 
cp.~Fig.~\ref{subleadingDiags}.
There are two important differences though: firstly, we include the worldline propagators~\eqref{WLprops}
on the outgoing line, unlike for the impulse where this line is cut.
Secondly, we leave the energy on the outgoing line finite ($\omega\neq0$)
unlike in the impulse where the outgoing energy is set to zero.
The Fourier transform on $\omega$ in Eq.~\eqref{trajFourier} yields $\omega=q\cdot v_1$,
and so there are now two mass-dimension scales present:
$|q|$, which is the overall momentum transfer, and $q\cdot v_1$, which is the energy on the outgoing line.

\section{Leading-Order Trajectory} \label{LeadingTrajectory}

Let us begin with the non-spinning case.
Without loss of generality, in the following we consider the trajectory of BH1.
At leading-PM order a single diagram contributes:
\begin{align}\label{leadingDeflection}
   z_1^{(1)\mu}(\omega)\,\,=\,\,
   \scalemath{.9}{\begin{tikzpicture}[baseline={25}]
      \begin{feynman}
      \vertex[dot] at ($(0,0)$) (a){};
      \vertex[dot] at ($(0,2)$) (b){};
      \vertex at ($(1.5,2)$) (c) {};
      \vertex at ($(-1.5,2)$) (d) {1};
      \vertex at ($(-1.5,0)$) (e) {2};
      \vertex at ($(1.5,0)$) (f) {};
      \diagram*{
      (a)--[dotted] (e),
      (a)--[dotted] (f),
      (a)--[photon, thick, momentum=\(q\)] (b),
      (b)--[dotted] (d),
      (b)--[fermion, edge label=\(z_1^\mu(\omega)\),thick](c),};
      \end{feynman}
      \end{tikzpicture}}\,.
\end{align}
Therefore, using the WQFT Feynman rules,
\begin{align}\label{z1fint}
z_1^{(1)\mu}(\tau_1)_{S^{0}}
&=-\i\frac{m_2 \kappa^2}{8}\int_q\frac{\deltabar(q\cdot v_2)}{q^2(q\cdot v_1+\i 0)^2} e^{\i q\cdot\tilde{b}(\tau_1)}
\\ & \qquad
[2(q\cdot v_1)(2\gamma v_2^\mu-v_1^\mu) -(2\gamma^2-1)q^\mu]\,.\nn
\end{align}
The Fourier transform on $\omega$~\eqref{ztauom} has here been performed,
yielding $\omega=q\cdot v_1$ and the shifted impact parameter $\tilde{b}^\mu(\tau_1)$~\eqref{shiftedB}.
Integrating out the external energy scale in this way is familiar from work on
the gravitational waveform~\cite{Jakobsen:2021lvp},
and yields the same kind of Fourier transforms on the total exchanged momentum $q^\mu$.
The $\i0$ prescription on the graviton is in this case immaterial.

To evaluate the leading trajectory we need two Fourier transform integrals both
belonging to the family
\begin{equation}\label{1PMfamily}
   I^{\mu_1\cdots\mu_n}_{\alpha,\beta}=
   \int_q \frac{\deltabar(q\cdot v_2)}{(q^2)^\alpha(q\cdot v_1+\i 0)^\beta}
   e^{\i q\cdot\tilde{b}(\tau_1)}\,q^{\mu_1}\cdots q^{\mu_n}\,.
\end{equation}
These are derived in Appendix~\ref{sec:1PMfourier}:
\begin{subequations}\label{ints1pm}
\begin{align}
I_{1,1}&
%=\int_{q}\frac{\deltabar(q\cdot v_2)}{q^2(q\cdot v_1+\i 0^{+})}\,e^{\i q\cdot \tilde{b}}
=\frac{\i }{4\pi\sqrt{\gamma^2-1}}\log(S(\tau))\Bigg\vert_{-\infty}^{\tau_1},\\
I^\mu_{1,2}&
%=\int_{q}\frac{\deltabar(q\cdot v_2)q^\mu}{q^2(q\cdot v_1+\i 0^{+})^2}\,e^{\i q\cdot 
%\tilde{b}}
= -\frac{\i}{4\pi\sqrt{\gamma^2-1}}\times
\label{Iintegral}\\ 
&\qquad\Bigg[\frac{b^{\mu}}{|b|^2}\bigg(\frac{D(\tau)}{\sqrt{\gamma^2-1}}
+
\tau\bigg) -w_1^\mu\log(S(\tau))\Bigg]\Bigg\vert_{-\infty}^{\tau_1}.\nn
\end{align}
\end{subequations}
We have adopted the notation of Ref~\cite{Bini:2024hme}:
\begin{subequations}
\begin{align}
D(\tau)&=\sqrt{|b|^2+(\gamma^2-1)\tau^2} \,,\\
S(\tau)&=\frac{\sqrt{\gamma^2-1}\,\tau+D(\tau)}{|b|}\, .
\end{align}
\end{subequations}
$D(\tau)$, referred to as $|\tilde{\mathbf{b}}|_2$ in \Rcite{Jakobsen:2021lvp},
measures the length of the shifted impact parameter in the rest frame of the second black hole.
As $S(\tau\to -\infty)\to 0$ we encounter a $\log(0)$ divergence
in our result for $\langle z_1^\mu(\tau_1)\rangle_{S^{0}}$.
%Though apparently problematic,
 As mentioned above, this is due to our choice of fixing the zeroth order parameters at $\tau_i=-\infty$. This 
%By using retarded propagators, $v_i^\mu$ and $b_i^\mu$ have been identified as parameters at $\tau_i=-\infty$.
 implies that the  black holes have already travelled an infinite distance once they reach a finite $\tau_i$, yielding a divergent constant. See e.g.~\cite{Bini:2024hme, Caron-Huot:2023vxl} for discussions on this intrinsic logarithmic divergence and its relation to the long-range effects of gravity. 

To obtain a finite result from the trajectory,
we subtract the same result evaluated at a finite reference time:
\begin{align}\label{z1wodiv}
& z_1^{(1) \mu}(\tau_1)_{S^0,\mathrm{ph}}= z_1^{(1)\mu}(\tau_1)_{S^0}- z_1^{(1)\mu}(0)_{S^0} \nn\\
&\qquad=m_2\frac{1}{(\gamma^2-1)^{3/2}}(\gamma(2\gamma^2-3)v_2^\mu+v_1^\mu)\log(S(\tau_1)) \nn\\
&\qquad\qquad+m_2(2\gamma^2-1)\frac{S(\tau_1)-1}{(\gamma^2-1)}\frac{b^\mu}{|b|}\,.
\end{align}
In fact, the trajectory is dependent on the frame choices and thus the conventions and boundary conditions one works with. Our trajectory is consistent with previous works in the WQFT 
formalism~\cite{Mogull:2020sak,Jakobsen:2021zvh,Jakobsen:2022psy,Jakobsen:2023oow,Haddad:2024ebn}.
While the authors of \Rcite{Bini:2022wrq} fix their boundary conditions differently, the
above result agrees with theirs
when acting on it with the projector $P^\mu_\nu=\delta_\nu^\mu-v_1^\mu v_{1,\nu}$. As further confirmation of our result we reproduce the 1PM momentum impulse:
\begin{align}\label{impCheck}
\begin{aligned}
\Delta p_1^{(1)\mu}&=m_1\left(\dot{z}_1^\mu(\tau_1\to\infty)-\dot{z}_1^\mu(\tau_1\to-\infty)\right)\\
&=m_1m_2\frac{(2\gamma^2-1)}{\sqrt{\gamma^2-1}}\frac{b^\mu}{|b|^2}+\cO(S)\,.
\end{aligned}
\end{align}
Finally, at this leading-PM order plugging the background field expansions~\eqref{bgexpansions}
into the on-shell condition~\eqref{onshellcond} for the non-spinning case, $g_{\mu\nu}\dot{x}_1^\mu\dot{x}_1^\nu=1$, gives
\begin{align}\label{leading-on-shell}
   0=G\,
   \Bigl (2v_1\cdot\dot{z}_1^{(1)}(\tau_1)+h_{\nu\rho}^{(1)}(b_1^\mu+\tau_1v_1^\mu)v_1^{\nu} v_1^{\rho}
   \Bigr)\,.
\end{align}
Unlike the corresponding on-shell condition on the momentum impulse $\Delta p_1^\mu$,
which at this perturbative order is simply $v_1\cdot\Delta p_1=0$,
here we also require the static contribution to the gravitational waveform
produced by the second black hole:
\begin{align}\label{leadingMetric}
\begin{aligned}
%  {h}_{\mu\nu}(k)  &=
\begin{tikzpicture}[baseline={(current bounding box.center)}]
   		\coordinate (inB) at (-1,-.5);
   		\coordinate (outB) at (1,-.5);
   		\coordinate (xB) at (0,-.5);
   		\coordinate (k) at (1,.5);
   		\draw (k) node [right] {${\mu,\nu}$};
   		\draw [dotted] (inB) -- (xB);
   		\draw [dotted] (xB) -- (outB);
   		\draw [fill] (xB) circle (.08);
   		\draw [graviton] (xB) -- (k) node [midway, above] {$k\quad$};
   		\draw (inB) node [left] {};
   		\draw [fill] (xB) circle (.08);
   		\end{tikzpicture}
\!\!\!=
\frac{\kappa m_2}{2k^2}e^{ik\cdot b_2}\dd(k\cdot v_2)\left(v_2^\mu v_2^\nu-\sfrac12\eta^{\mu\nu}\right).
\end{aligned}
\end{align}
% Stripping off the overall $e^{ik\cdot b_2}\dd(k\cdot v_2)$,
% the waveform may be written as
% \begin{equation}
%    h_{\mu\nu}(x_1(\tau))=\int_k e^{\i k\cdot \tilde{b}(\tau_1)}\deltabar(k\cdot v_2) {h}_{\mu\nu}(k)+\cO(\kappa^3)\, ,
% \end{equation}
%\begin{equation}\label{fourierh}
%h_{\mu\nu}(x_1(\tau))=\int_k e^{-\i k\cdot(b_1+\tau v_1)}\tilde{h}_{\mu\nu}(k)+\cO(\kappa^3)\,.
%\end{equation}
% {\color{violet}
%  where we defined ${h}_{\mu\nu}(k)$ via
% \gustav{I will implement my approach!}
% \begin{equation}
%    e^{i k\cdot b_2}\deltabar(k\cdot v_2){h}_{\mu\nu}(k):={h}_{\mu\nu}(k)\, .
% \end{equation}} 
%In this notation we have
%\begin{equation}
  % h_{\mu\nu}(x_1(\tau))=\int_k e^{\i k\cdot \tilde{b}}\deltabar(k\cdot v_2)h_{\mu\nu}(k)+\cO(\kappa^3)\, ,
%\end{equation}
% aligning with the conventions we have introduced in \eqref{trajFourier}.
Putting together the pieces,
\begin{align}
   2v_1\cdot\dot{z}_{1}^{(1)}(\tau_1)=
   - h_{\mu\nu}^{(1)}(b_1^\mu+\tau_1v_1^\mu)v_1^\mu v_1^\nu
   =\frac{2m_2(2\gamma^2-1)}{D(\tau_1)}\,,
\end{align}
thus verifying the consistency requirement~\eqref{leading-on-shell}.
%\begin{align}
%\begin{aligned}
%&\alpha^{\prime\, \mu}_1=\frac{G m_2}{|b|^2(\gamma^2-1)D(\tau)} [2\gamma(\gamma^2-1) (a_{-\infty,\, 1}\cdot v_2)\tau b^\mu\\&\qquad+|b|\gamma (|b|-D(\tau)) (a_{-\infty,\, 1}\cdot v_2)v_1^\mu\\
%&\qquad\qquad+(\gamma^2-1)(b\cdot \alpha_{-\infty,\, 1})\tau(v_1^\mu-2\gamma v_2^\mu)]\ .
%\end{aligned}
%\end{align}

Moving on to the spinning case, we compute the trajectory up to quadratic order in spins.
For compactness, we present here the results only up to linear order in spins ---
for the full quadratic order, we refer the reader to our \zenodo
repository submission.
With a single diagram contributing,
the main difference to the non-spinning case is that we require
higher-rank integrals of the family~$I^{\mu_1\cdots\mu_n}_{\alpha,\beta}$~\eqref{1PMfamily},
%\begin{equation}
%I^{\mu_1...\mu_n}_{\alpha,\, l}=
%\int_q\frac{e^{\i q\cdot \tilde{b}}\deltabar(q\cdot v_2)q^{\mu_1}...q^{\mu_n}}{(q^2)^\alpha(q\cdot v_1+\i 0^{+})^l}\,,
%end{equation}
up to $n=3$.
As described in Appendix~\ref{sec:1PMfourier},
these integrals are obtained as $b$-derivatives of the scalar integrals with the same powers of the propagators.
%These integrals are obtained as higher derivatives of the scalar integral $J$~\eqref{ints1pm}.
% Moving on to the spinning contributions
% the spinning Feynman-rules yields the deflection up to quadratic order one has\gustav{has the divergent part already been subtracted?}
% \begin{equation}
% \begin{split}
% &\langle z_{1}^\mu(\tau)_{S^1,S^2}\rangle_{\text{ph}}=\frac{\kappa ^2 m_2}{8}\int_q e^{iq\cdot b-q\cdot v_1\tau}\frac{\deltabar(q\cdot v_2)}{q^2 (q\cdot v_{1} +\i 0^{+})^2}\Big[\big(2 \gamma  q^{\mu } \big[(q\cdot S_{1}\cdot v_2)-(q\cdot S_2\cdot v_1)\big]\\
% &+q\cdot v_1 
%    \big(
%    2 v_2^{\mu } \big[(q\cdot S_{2}\cdot v_1)-(q\cdot S_1\cdot v_2)\big]+(q\cdot S_1)^\mu+2 \gamma 
%   (q\cdot S_{2})^\mu\big)\big)\\
%   %
%   &+i \big(2 q^{\mu } \big[2 (q\cdot S_{1}\cdot v_2)(q\cdot S_2\cdot v_1)-2
%    (q\cdot S_{1}\cdot v_2)^2- (q\cdot S_1)_{\lambda} (S_1\cdot q)^{\lambda}\\
%    &-2 \gamma  (q\cdot S_1)_{\lambda}(S_2\cdot q)^{\lambda}
%    -2
%    (q\cdot S_{2}\cdot v_1)^2-(q\cdot S_2)_{\lambda}(S_2\cdot q)^{\lambda}\big]+ q\cdot v_1  (v_1^{\mu } (q\cdot S_2)_{\lambda}(S_2\cdot q)^{\lambda}\\
%    &+ v_2^{\mu } (q\cdot S_{1})_{\lambda}(S_2\cdot q)^{\lambda})
%    %
%    - q\cdot v_1 (q\cdot S_{2})^{\mu}
%    \big[(q\cdot S_{1}\cdot v_2)-2 (q\cdot S_2\cdot v_1)\big]\big)\Big] .
%    \end{split}
% \end{equation}
The 1PM linear-in-spin trajectory is then given by
% , with 
% $l^\mu=-\eps^\mu_{\nu\rho\sigma}b^\nu v_1^\rho v_2^\sigma$,
% where we also introduce a hat symbol to denote the unit normalized vectors, such that for a vector $w$, $\hat{w}:=w/|w|$.
\begin{align}
&z_1^{(1)\mu}(\tau_1)_{S^1, \mathrm{ph}}\\
&=-\frac{m_2}{\hat{D}(\tau_1)|b|}
\bigg[\gamma\, \bar{\tau}_1 \, w_1^\mu a_1\cdot \hat{l}
-\bar{\tau}_1\, v_2^\mu (a_1\cdot \hat{l}+2\, a_2\cdot \hat{l})\nn\\
&\quad-\frac{\hat{b}^\mu}{v} \Big[ \Big(\hat{D}(\tau_1) (2 
\bar{\tau}_1-1)+2 \bar{\tau}_1^2+1 \Big) a_1\cdot \hat{l}\nn\\
&\qquad\quad+2 \Big(\hat{D}(\tau_1) (\bar{\tau}_1-1)+\bar{\tau}_1^2+1 \Big) a_2\cdot 
\hat{l}\Big]\nn\\
&\quad-\frac{ \hat{l}^\mu}{(\gamma v)^3} \Big[\bar{\tau}_1 (2 \gamma \Big((\gamma v)^2 
(\hat{D}(\tau_1)+\bar{\tau}_1) a_2\cdot 
\hat{b}+\gamma v a_2\cdot v_1)\nn\\
&-\gamma v 
a_1\cdot v_2\Big)+\gamma  (\gamma v)^2 \Big(\hat{D}(\tau_1) (2 
\bar{\tau}_1\!-\!1)+2 \bar{\tau}_1^2+1\Big)\, a_1\cdot \hat{b}\Big]\bigg],\nn
\end{align}
where we have used
\begin{align}\label{Dtau}
\bar{\tau}_1&:=\frac{\tau_1}{|b|\gamma v}\,,& \hat{D}(\tau_1)&:=\frac{D(\tau_1)}{|b|}\, .
\end{align}
%Unlike the non-spinning part of the same result~\eqref{z1wodiv}
%there is no dependence on $\log(S(\tau_1))$.
The spinning correction to the trajectory is finite since spin effects do not affect the gravitational long-range interactions which cause the logarithmic drift in the trajectory discussed above.
As in the non-spinning case~\eqref{impCheck},
we checked this result by confirming that it reproduces the
momentum impulse in the asymptotic limit,
an expression which may be found in Ref.~\cite{Jakobsen:2021zvh}.
We also checked the on-shell condition~\eqref{onshellcond}, which at leading-PM order becomes
\begin{align}\label{quadSpinRelation}
\begin{aligned}
   &2v_1\cdot\dot{z}^{(1)}_1(\tau_1)+h^{(1)}_{\mu\nu}(b_1^\mu+\tau_1v_1^\mu)v_1^\mu v_1^\nu\\
   &\qquad=
   \frac1{2m_1^2}\partial_\nu\partial_\sigma h^{(1)}_{\mu\rho}(b_1^\mu+\tau_1v_1^\mu)S_{-\infty,1}^{\mu\nu}S_{-\infty,1}^{\rho\sigma}\,.
\end{aligned}
\end{align}
Here we have used $R_{\mu\nu\rho\sigma}=\sfrac\kappa2(\partial_\nu\partial_\rho
 h_{\mu\sigma}+\partial_\mu\partial_\sigma h_{\nu\rho}-\partial_\mu\partial_\rho
  h_{\nu\sigma}-\partial_\nu\partial_\sigma h_{\mu\rho})+\cO(\kappa^2)$.
While it can be ignored at linear order in spin,
at quadratic order in spin the Riemann curvature tensor enters the calculation.
We confirm that Eq.~\eqref{quadSpinRelation}
holds at every point in time $\tau_1$. 

In order to obtain the leading-order spin deflection $a_1^{(1)\mu}(\tau_1)$,
we first derived an expression for $\alpha_1^{(1)\mu}(\tau_1)$.
Like the deflection $z_1^{(1)\mu}(\omega)$~\eqref{leadingDeflection},
the corresponding frequency-space result is given by a single diagram:
\begin{align}
   \alpha_1^{(1)\mu}(\omega)\,\,=\,\,
   \scalemath{.9}{\begin{tikzpicture}[baseline={25}]
      \begin{feynman}
      \vertex[dot] at ($(0,0)$) (a){};
      \vertex[dot] at ($(0,2)$) (b){};
      \vertex at ($(1.5,2)$) (c) {};
      \vertex at ($(-1.5,2)$) (d) {1};
      \vertex at ($(-1.5,0)$) (e) {2};
      \vertex at ($(1.5,0)$) (f) {};
      \diagram*{
      (a)--[dotted] (e),
      (a)--[dotted] (f),
      (a)--[photon, thick, momentum=\(q\)] (b),
      (b)--[dotted] (d),
      (b)--[fermion, edge label=\(\alpha_1^\mu(\omega)\),thick](c),};
      \end{feynman}
      \end{tikzpicture}}\,.
\end{align}
We have checked that the charge $\alpha_1\cdot \bar{\alpha}_1$ is conserved
and that the spin-supplementary condition
(SSC) $S_{i}^{\mu\nu}\dot{x}_{i,\nu}=0$ holds at all times,
which is implied by $\alpha_i\cdot\dot{x}_i=0$.
Using Eqs.~\eqref{spintensor}, \eqref{SVec},
we are consequently able to obtain the leading-order change in the Pauli-Lubanski spin vector:
\begin{align}
&{a_1^{(1)\mu}(\tau_1)}_{S^1}=\\
&\frac{m_2}{\hat{D}(\tau_1)|b|}\bigg[-2 (a_1\cdot \hat{l}) \hat{l}^\mu-2  (a_1\cdot b-\gamma  \hat{\tau}_1 a_1\cdot \
v_2)\hat{b}^\mu\nn\\
&- (a_1\cdot b\, ((1-2 \gamma ^2) \gamma  v
\hat{D}(\tau_1)+(\gamma v)^2 \hat{\tau}_1 )+3 \gamma  a_1\cdot 
v_2)w_1^\mu\nn\\
&-\frac{ (a_1\cdot b\, ((2 \gamma ^2-1)  
\hat{D}(\tau_1) v+\gamma \hat{\tau}_1 v^2)- a_1\cdot 
v_2)v_2^\mu}{v^2}\bigg]\, ,\nn
\end{align}
where $\hat{\tau}_1:=\tau_1/|b|$.

%\begin{align}
%\begin{aligned}
%\label{riempart}
%&2 v_1\cdot \dot{z_1}=\frac{G m_2 }{D(\tau)^{5}}\big[a_1^2 \big(\big(\gamma ^2-1\big) \tau ^2+\big(3 \gamma ^2-2\big)
   %|b|^2\big)\\
   %&+\big(6 \gamma ^2-3\big) (a_1\cdot b)^2+6 \gamma  \tau  (a_1\cdot b)
  % (a_1\cdot v_2)\\
  % &+3 (|b|^2+\tau^2) (a_1\cdot v_2)^2+\big(\big(-2 \gamma ^4+5 \gamma ^2-3\big) \tau ^2\\
   %&+\big(4 \gamma ^2-3\big) |b|^2\big) S_2^2+6 \tau  (S_2\cdot b) (S_2\cdot
   %v_1)\\
  % &-3(2\gamma^2-1) (S_2\cdot b)^2-(4|b|^2+2\gamma^2\tau^2-5\tau^2) (S_2\cdot v_1)^2\big]\ .
%\end{aligned}
%\end{align}

%\begin{align}
%\begin{aligned}
%\label{riempart2}
%&\kappa h_{\mu\nu}v_1^\mu v_1^\nu =-\frac{G m_2 }{D(\tau)^{5}}\big[\big(\big(-2 \gamma ^4+5 \gamma ^2-3\big) \tau ^2\\
%&+\big(4 \gamma ^2-3\big) |b|^2\big) S_2^2+6 \tau  (S_2\cdot b) (S_2\cdot
  % v_1)\\
   %&-3(2\gamma^2-1) (S_2\cdot b)^2-(4|b|^2+2\gamma^2\tau^2-5\tau^2) (S_2\cdot v_1)^2\big]\ .
%\end{aligned}
%\end{align}

%\begin{align}
%\begin{aligned}
%&2 v_1\cdot \dot{z_1}+\kappa h_{\mu\nu}v_1^\mu v_1^\nu =R_{abcd}\left[\bar{\alpha}_{1}^{a}\alpha_{1}^{b}\bar{\alpha}_{1}^{c}\alpha_{1}^{d}\right]_{-\infty}\\
%&=\frac{G m_2 }{D(\tau)^{5}}\big[a_1^2 \big(\big(\gamma ^2-1\big) \tau ^2+\big(3 \gamma ^2-2\big)
   %|b|^2\big)\\
   %&+\big(6 \gamma ^2-3\big) (a_1\cdot b)^2\\
   %&+6 \gamma  \tau  (a_1\cdot b)
   %(a_1\cdot v_2)+3 (|b|^2+\tau^2) (a_1\cdot v_2)^2\big]\ .
%\end{aligned}
%\end{align}

\section{Sub-leading Trajectory}

The sub-leading trajectory $\cZ_1^{(2)\mu}$ comprises two mass sectors
which are proportional to $m_2^2$ and $m_1 m_2$.
All contributing diagrams are displayed in Fig.~\ref{subleadingDiags},
with terms carrying $m_2^2$ contributing in the probe limit $m_1\ll m_2$.
Just as at 1PM order, we perform the trivial Fourier transform integral on the outgoing energy $\omega$,
which introduces the shifted impact parameter
$\tilde{b}^\mu(\tau_1)=b^\mu-v_1^\mu\tau_1$ in the exponential
of the $q$-Fourier transform~\eqref{trajFourier}.
Unlike at 1PM order though, and given the complexity involved,
we will not perform this Fourier transform explicitly;
instead, $\cZ_1^{(2)\mu}(q)$ is the focus of this section.
Our object of interest will ultimately be the total radiated angular momentum from the system $J_{\rm rad}^\mu$,
which is negative the asymptotic change in total angular momentum of the system $\Delta J^\mu$.
In Section~\ref{sec:jRad} we will show how the latter is directly extracted from $\cZ_i^{\mu}(q)$ ---
performing the Fourier integrals in the simpler asymptotic limit $\tau_i\to\pm\infty$.

\subsection{Integration}

% At $N$PM order, $z_i^\mu(q)$ can be written as a linear combination of Fourier transforms
% of $N-1$-loop integrals of the form
% \begin{align}
% \begin{aligned}
%    \int_q e^{\i q\cdot \tilde{b}} \deltabar(q\cdot v_2)I_{N-1\mathrm{-loop}}\, .
% \end{aligned}
% \end{align}

The 2PM trajectory $\cZ_1^{(2)\mu}$
comprises one-loop integrals of two different families,
emerging from the two mass sectors $m_2^2$ and $m_1m_2$.
The integral families, which are always multiplied by an overall $\deltabar(q\cdot v_{2})$, are
\begin{subequations}
\begin{align}\label{probeFamily}
   &I^{1-\mathrm{loop}}_{\nu_1,\nu_2,\nu_3}(|q|, q\cdot v_1, \gamma)=\\
   &\int_\ell\frac{\deltabar(\ell\cdot v_2)}{(\ell^2)^{\nu_1}((\ell-q)^2
   )^{\nu_2}(\ell\cdot v_1+\i 0)^{\nu_3}}\, ,\nn
\end{align}
corresponding to the $m_2^2$ terms, and 
\begin{align}
   &J^{1-\mathrm{loop}}_{\nu_1,\nu_2,\nu_3}(|q|, q\cdot v_1, \gamma)=\\
   &\int_\ell\frac{\deltabar(\ell\cdot v_1)}{(\ell^2)^{\nu_1}((\ell-q)^2+\i 0\ \mathrm{sgn}(\ell^0-q^0))^{\nu_2}(\ell\cdot v_2+\i 0)^{\nu_3}}\, ,\nn
\end{align}
\end{subequations}
corresponding to the $m_1m_2$ terms.
We stress that, due to the suppressed multiplication with $\deltabar(q\cdot v_2)$,
these integrals are  not related by symmetry and depend on \emph{two} dimensionful parameters:
$|q|$ and $q\cdot v_1$.x
This is  unlike the single-scale integrals emerging in the calculation of the momentum impulse that cannot depend
on $q\cdot v_1$ due to the presence of  $\deltabar(q\cdot v_1)$.
Here we may still assume that $q\cdot v_2=0$, which leads to simplifications of the loop integrands. Interestingly, the integral families form a subset of the integral families arising in the one-loop gravitational waveform \cite{Caron-Huot:2023vxl, Bohnenblust:2023qmy, Brandhuber:2023hhy, Herderschee:2023fxh}.

While a complete account of the loop integration is contained in Appendix~\ref{loopIntegration},
we remark here on some important features.
By introducing $\hat{q}\cdot v_1=q\cdot v_1/|q|$
the overall mass dimension is relegated to overall powers of $|q|$,
and thus factored from the integrals.
All non-trivial functional dependence is therefore contained in $\gamma$ and $\hat{q}\cdot v_1$.
In the probe-limit family~\eqref{probeFamily} the dependence on $\gamma$
can be established by entering the frame $v_2^\mu=(1,\mathbf{0})$.
In both families, dependence on $\hat{q}\cdot v_1$ is fixed by the method of differential equations (DEs),
considering also integration-by-parts identities (IBPs) between integrals of the same family. 
Boundary conditions on these integrals are provided by the limit $\hat{q}\cdot v_1\to0$. In particular, we keep the full dependence on $\gamma$ in the boundary integrals.
The $\omega\to 0$ limit comprises two integration regions (after a shift $\ell\to \ell+q$):
\begin{equation}
   \ell^\mu \sim (\hat q\cdot v_1, 1)\ , \qquad \ell^\mu \sim (\hat q\cdot v_1,\hat q\cdot v_1)\,.
\end{equation}
The first of these produces boundary integrals that are familiar from one-loop computations of asymptotic quantities such as the momentum impulse $\Delta p_1^\mu$.
The other is a new region that requires separate evaluation, 
and we comment on it in Appendix~\ref{loopIntegration}. 
Note that, despite the apparent similarity,
these regions are entirely unrelated to the potential and radiative regions
used when fixing $\gamma$-dependent integrals in the limit $v\to0$ ---
both regions contribute to both mass sectors.

% {\color{violet}In contrast to considering radiative
%  versus potential regions, we do not take the $\gamma\to 1$ limit on the boundary.
%   Both mass sectors get contributions from the new region, while only
%    the mass sector proportional to $m_1 m_2$ has a non-vanishing radiative region.}

To further illustrate the structure of the integrals,
we present the master integrals of each family.
Firstly, the probe-limit family comprises three master integrals (normalized by a factor of $(4\pi)^{-2\eps}e^{\gamma_{\mathrm{E}}\eps}$):
% where we have introduced the relative velocity $v=\sqrt{\gamma^2-1}/\gamma$:
%
\begin{subequations}\label{eq:1loopI}
   \begin{align}
   &I^{1-\mathrm{loop}}_{0,1,1}=\frac{\i|q|^{-2\epsilon}}{8 \pi\gamma v}\bigg(\frac{1}{\eps}+\i\pi+2\log[(\hat{q}\cdot v_1)_N^{-1}]\bigg)\nn\\
   &\quad+\mathcal{O}(\eps)\, ,\\
   &I^{1-\mathrm{loop}}_{1,1,0}=\frac{1}{8}|q|^{-1-2\epsilon}+\mathcal{O}(\eps)\, ,\\
& I^{1-\mathrm{loop}}_{1,1,1}=\frac{\i|q|^{-2\epsilon}}{8 \pi |q|^2\gamma v}\bigg(\frac{1}{\eps}-\i\pi-2\log[(\hat{q}\cdot v_1)^{-1}_N]\bigg)
\\
& +\mathcal{O}(\eps)\, ,\nn
 \end{align}
\end{subequations}

where we defined $(\hat{q}\cdot v_1)_N:=\hat{q}\cdot v_1/(\gamma v)$.
Meanwhile, the four masters of integral
family $J^{1-\mathrm{loop}}_{\nu_1,\nu_2,\nu_3}$ are

\begin{subequations}\label{eq:1loopJ}
   \begin{align}
&J^{1-\mathrm{loop}}_{0,1,0}=\i\,|q|^{1-2\epsilon}\,\frac{\hat{q}\cdot v_1}{4 \pi}+\mathcal{O}(\eps)\, ,\\
&J^{1-\mathrm{loop}}_{0,1,1}=-\frac{\i|q|^{-2\epsilon}}{8\pi\gamma v}\bigg(\frac{1}{\eps}+\i\pi+2\arccosh(\gamma)\\
&\quad+2\log[(\hat{q}\cdot v_1)^{-1}_N]\bigg)+\mathcal{O}(\eps)\, ,\nn\\
&J^{1-\mathrm{loop}}_{1,1,0}=-\frac{\i\, |q|^{-1-2\eps}}{8\pi\sqrt{(\hat{q}\cdot v_1)^2+1}}\\
&\quad \times\Bigg(\i\pi+ 2\,
 \mathrm{arctanh}\bigg(\frac{\hat{q}\cdot v_1}{\sqrt{(\hat{q}\cdot 
v_1)^2+1}}\bigg)\Bigg)+\mathcal{O}(\eps)\,,\nn\\
 &J^{1-\mathrm{loop}}_{1,1,1}=\frac{\i|q|^{-2\epsilon}}{8\pi|q|^2\gamma v}\bigg(-\frac{1}{\eps}+\i\pi+2\arccosh(\gamma)\\
 &\quad+2\log[(\hat{q}\cdot v_1)_N^{-1}]\bigg)+\mathcal{O}(\eps)\, .\nn
   \end{align}
\end{subequations}
Notice the appearance in these integrals of $\arccosh(\gamma)$,
which in a calculation of e.g.~the momentum impulse shows up only in two-loop integrals ---
for us here, it will ultimately contribute to the radiated angular momentum.
\subsection{Frequency-domain trajectory} 

Evaluation of the diagrams in Fig.~\ref{subleadingDiags} yields
the 2PM trajectory in $q$-space.
This comprises results for $z_1^{(2)\mu}(q)$ and $\alpha_1^{(2)\mu}(q)$ ---
and consequently $a_1^{(2)\mu}(q)$ --- up to linear order in spin.
In order to illustrate our results here we present the non-spinning part
of $z_1^{(2)\mu}(q)$ only, with linear-in-spin provided in our \zenodo submission.
Our result is split into two individually gauge-invariant mass sectors:
\begin{align}\label{subleadingTr}
\begin{aligned}
 &z_{1}^{(2)\mu}(q)=C^{(m_{2}^{2})}_{1} w_1^\mu+C^{(m_{2}^{2})}_{2} v_2^\mu+C^{(m_{2}^{2})}_{3}\hat{q}^\mu\\
& +C^{(m_{1}m_{2})}_{1} w_1^\mu+C^{(m_{1}m_{2})}_{2} v_2^\mu+C^{(m_{1}m_{2})}_{3}\hat{q}^\mu\ ,
 \end{aligned}
\end{align}
with
\begin{widetext}
\begin{subequations}
\begin{flalign}
\begin{aligned}
  &C_{1}^{(m_2^2)}=
-\frac{ \pi m_2^2 |q|^{-2-2 \epsilon}}{\gamma v (\hat{q}\cdot v_1)^2}\bigg[\frac{2 ((1-2 \gamma 
^2)^2+2 (\hat{q}\cdot v_1)^2)}{ \epsilon  
}\\
&+ (-8 \gamma ^2-3 \i 
\pi  (\gamma v)^{3} \hat{q}\cdot v_1+4 (\hat{q}\cdot 
v_1)^2 (2\gamma ^2-\i\pi-2\log [(\hat{q}\cdot v_1)^{-1}_N])+4)\bigg]+\cO(\eps)\, ,   
\end{aligned}&&
\end{flalign}
\begin{flalign}
\begin{aligned}
   &C_{2}^{(m_2^2)}=-\frac{4 \pi  \gamma  m_2^2 |q|^{-2-2 \epsilon}}{(\gamma v)^{3}}\bigg[\frac{1}{
 \epsilon }+\frac{ ( \hat{q}\cdot
v_1 (2\gamma ^2-\i\pi-2\log [(\hat{q}\cdot v_1)_N^{-1}]-2)-3 \i \pi  (\gamma v)^{3})}{ \hat{q}\cdot v_1}\bigg]
+\cO(\eps)\, ,
\end{aligned}&&
\end{flalign}
\begin{flalign}
\begin{aligned}
   &C_{3}^{(m_2^2)}=-\frac{\pi m_2^2|q|^{-2-2\epsilon} }{(\gamma v)^3\hat{q}\cdot v_1}\bigg[\frac{2  (1-2 \gamma ^2)}{
 \epsilon } +\frac{1}{2 (\hat{q}\cdot v_1)} \Big(3 \i \pi  
\gamma v (5 \gamma ^4-6 \gamma ^2+1)-7 \i \pi  (\gamma 
v)^{3} (\hat{q}\cdot v_1)^2-4 \hat{q}\cdot v_1 (4 (\gamma 
v)^2 \gamma ^2\\
&-(2\gamma^2-1)(\i\pi+2\log [(\hat{q}\cdot v_1)_N^{-1}]))\Big)\bigg]+\cO(\eps)\, .
\end{aligned}&&
\end{flalign}
\begin{flalign}
\begin{aligned}
&C_{1}^{(m_1m_2)}=\pi m_1 m_2 |q|^{-2-2\epsilon}\bigg[
-\frac{2 \gamma ((1-2 \
\gamma ^2)^2+(6-4 \gamma ^2) (\hat{q}\cdot \
v_1)^2)}{\gamma v \epsilon  (\hat{q}\cdot v_1)^2}\\
&+\frac{1}{6} \bigg(-\frac{24 
\gamma  (2 \gamma ^2-3) (2\,\mathrm{arccosh}(\gamma )+\i\pi +2\log[(\hat{q}\cdot v_1)_N^{-1}])}{\gamma 
v}-\frac{24 \gamma  (-2 \gamma ^2+2 \gamma ^2 (\hat{q}\cdot
v_1)^2+1)}{\gamma v (\hat{q}\cdot v_1)^2}\\
&+\frac{2 (61
\gamma ^4-80 \gamma ^2+24 \gamma ^2 (2 \gamma ^2-3) 
(\hat{q}\cdot v_1)^6+2 (70 \gamma ^4-105 \gamma ^2-1) 
(\hat{q}\cdot v_1)^4+(168 \gamma ^4-233 \gamma ^2-7) 
(\hat{q}\cdot v_1)^2-5)}{((\hat{q}\cdot v_1)^2+1)^3}\\
&-\frac{3\Big( \i 
\pi+2\,\mathrm{arctanh}\Big(\frac{\hat{q}\cdot 
v_1}{\sqrt{(\hat{q}\cdot v_1)^2+1}}\Big)\Big)}{\hat{q}\cdot v_1 ((\hat{q}\cdot 
v_1)^2+1)^{7/2}}  \Big(3 (5 \gamma ^4-8 \gamma ^2+3)+8 (2 \gamma ^4-3 \gamma 
^2+1) (\hat{q}\cdot v_1)^6+8 (7 \gamma ^4-10 \gamma ^2+3) 
(\hat{q}\cdot v_1)^4\\
&+25 (2 \gamma ^4-3 \gamma ^2+1) 
(\hat{q}\cdot v_1)^2\Big)\bigg)\bigg]
+\cO(\eps)\, ,
\end{aligned}&&
\end{flalign}
\begin{flalign}
\begin{aligned}
&C_{2}^{(m_1 m_2)}=\pi m_1 m_2 |q|^{-2-2\epsilon}\bigg[\frac{2  ((\gamma v)^2 (1-2 
\gamma ^2)^2+(4 \gamma ^4-6 \gamma ^2) (\hat{q}\cdot v_1)^2)}{(\gamma 
v)^{3} \epsilon  (\hat{q}\cdot v_1)^2}\\
&+\frac{1}{6}  \bigg(-\frac{24
(2 \gamma ^2-3) \gamma ^2(2\,\mathrm{arccosh}(\gamma )+\i\pi+2\log[(\hat{q}\cdot v_1)_N^{-1}])}{(\gamma 
v)^{3}}-\frac{24 (2 \gamma ^2+6 \gamma ^2 (\hat{q}\cdot 
v_1)^2-1)}{\gamma v (\hat{q}\cdot v_1)^2}\\
&+\frac{3 \Big(\i \pi +2\,\mathrm{arctanh}\Big(\frac{\hat{q}\cdot 
v_1}{\sqrt{(\hat{q}\cdot v_1)^2+1}}\Big)\Big) \gamma  (3 (5 \gamma ^2+7)+8 (2 
\gamma ^2+1) (\hat{q}\cdot v_1)^6+8 (7 \gamma ^2+5) 
(\hat{q}\cdot v_1)^4+(50 \gamma ^2+53) (\hat{q}\cdot 
v_1)^2)}{\hat{q}\cdot v_1 ((\hat{q}\cdot 
v_1)^2+1)^{7/2}}\\
&-\frac{2 \gamma  (61 \gamma ^4-70 
\gamma ^2+24 \gamma ^2 (2 \gamma ^2-1) (\hat{q}\cdot v_1)^6+2 
(70 \gamma ^4-55 \gamma ^2+21) (\hat{q}\cdot v_1)^4+3 (56 \gamma 
^4-57 \gamma ^2+25) (\hat{q}\cdot v_1)^2+33)}{(\gamma v)^2 
((\hat{q}\cdot v_1)^2+1)^3}\bigg)\bigg]\\
&+\cO(\eps)\, ,
\end{aligned}&&
\end{flalign}
\begin{flalign}
\begin{aligned}
&C_{3}^{(m_1 m_2)}=\frac{\pi m_1 m_2 |q|^{-2-2\epsilon}}{\hat{q}\cdot v_1}\bigg[
\frac{2  \gamma  (-4 \gamma ^4+8 \gamma ^2-3) 
}{(\gamma v)^{3} \epsilon  
}\\
&+\frac{1}{6 
(\hat{q}\cdot v_1)} \bigg(\frac{12 (4 \gamma ^4-8 \gamma 
^2+3) \gamma (2\,\mathrm{arccosh}(\gamma )+\i\pi+2\log[(\hat{q}\cdot v_1)_N^{-1}]) \hat{q}\cdot v_1}{(\gamma 
v)^{3}}+\frac{96 \gamma ^3 \hat{q}\cdot v_1}{\gamma 
v}\\
&-\frac{3 \Big(\i \pi+2\, \mathrm{arctanh}\Big(\frac{\hat{q}\cdot v_1}{\sqrt{(\hat{q}\cdot 
v_1)^2+1}}\Big) \Big)
(15 \gamma ^2+16 \gamma ^2 (\hat{q}\cdot v_1)^6+(52 \gamma ^2-2) 
(\hat{q}\cdot v_1)^4+(46 \gamma ^2-5) (\hat{q}\cdot 
v_1)^2-3)}{((\hat{q}\cdot v_1)^2+1)^{7/2}}\\
&+\frac{2 
\hat{q}\cdot v_1 (5 \gamma ^4+30 \gamma ^2+12 (2 \gamma ^2-1) 
(\hat{q}\cdot v_1)^6+(80 \gamma ^2-44) (\hat{q}\cdot v_1)^4+(20 
\gamma ^4+71 \gamma ^2-55) (\hat{q}\cdot v_1)^2-23)}{(\gamma 
v)^2 ((\hat{q}\cdot v_1)^2+1)^3}\bigg)\bigg]
+\cO(\eps)\, .
\end{aligned}&&
\end{flalign}
\end{subequations}
\end{widetext}
The $1/\epsilon$ poles which are present in $z^\mu_1(q)$ either
cancel or yield finite terms after Fourier integration in all asymptotic observables that we have computed. Thus they do not affect our argument that the trajectory is a suitable tool to efficiently derive asymptotic observables, in particular the radiated angular momentum.
 Analogously we found the spin trajectory which is made use of in later checks.
 However, we do not display it but refer to our  \zenodo repository
 submission.

\subsection{Checks}

We performed two main checks on the sub-leading trajectory~\eqref{subleadingTr}.
Firstly, as at leading order, we confirmed that it reproduces the momentum impulse $\Delta p_1^\mu$ in the asymptotic limit $\tau_1\to\pm\infty$,
wherein the $q$-Fourier transform simplifies due to $q\cdot v_1=0$.
We have
\begin{align}
\label{limitfordp}
\begin{aligned}
   \Delta p_1^\mu&=m_1\int_{-\infty}^\infty\d\tau_1\frac{\d^2 z_1^\mu(\tau_1)}{\d\tau_1^2}\\
   &=m_1\int_qe^{iq\cdot b}\dd(q\cdot v_1)\dd(q\cdot v_2)(-iq\cdot v_1)^2z_1^\mu(q)\,,
\end{aligned}
\end{align}
having inserted Eq.~\eqref{trajFourier} and performed the $\tau_1$-integration.
Inserting the sub-leading trajectory~\eqref{subleadingTr},
the Fourier integrals are well known, 
and we reach agreement with the literature ---
see \Rcite{Jakobsen:2021zvh} both for the Fourier transform and momentum impulse.
In a similar manner, using the time-dependent spin trajectory
we have also reproduced the 2PM spin kick~\cite{Jakobsen:2021zvh}.

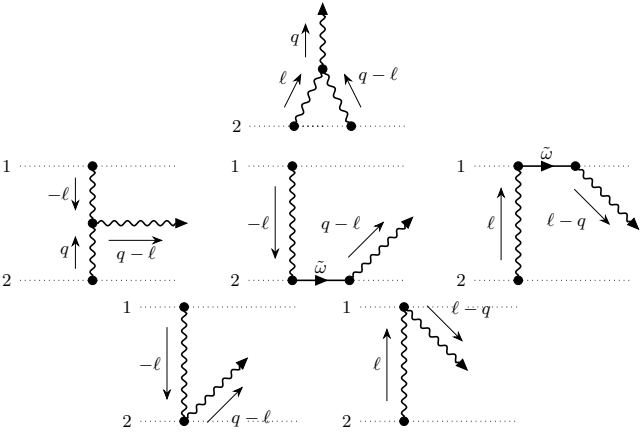
\begin{figure}[t!]
\begin{center}
   \scalemath{0.76}{
\begin{tikzpicture}
\begin{feynman}
\vertex at ($(0,0)$) (a);
\vertex at ($(1.5,0)$) (b);
\vertex[dot] at ($(-0.5,0)$) (c){};
\vertex[dot] at ($(0.5,0)$) (d){};
\vertex[dot] at ($(0,1)$) (e){};
\vertex [isosceles triangle, draw, rotate=90, fill=black, scale=0.4] at ($(0,2)$) (f) {};
\vertex at ($(-1.5,2)$) (g);
\vertex at ($(1.5,2)$) (h);
\vertex at (-1.5,0) (i) {2};
\diagram*{
a--[dotted] (c),
c--[dotted] (d),
d--[dotted] (b),
c--[photon, thick, momentum=\(\ell\)] (e),
a--[dotted](i),
e--[photon, thick, rmomentum=\(q-\ell\)] (d),
e--[photon, thick, momentum=\(q\)] (f),
%f--[scalar] (g),
};
\end{feynman}
\end{tikzpicture}}
   \scalemath{0.76}{
   \begin{tikzpicture}
   \begin{feynman}
   \vertex[dot] at ($(0,0)$) (a){};
   \vertex at ($(1.5,0)$) (b);
   \vertex at ($(-0.5,0)$) (c);
   \vertex at ($(0.5,0)$) (d);
   \vertex[dot] at ($(0,1)$) (e){};
   \vertex at ($(-1.5,2)$) (g) {1};
   \vertex at ($(1.5,2)$) (h);
   \vertex at (-1.5,0) (i) {2};
   \vertex[dot] at ($(0,2)$) (j){};
   \vertex[isosceles triangle, draw, fill=black, scale=0.4 ] at ($(1.5,1)$) (k) {};
   \diagram*{
   %(a)--[scalar] (c),
   (a)--[photon,thick,momentum=\(q\)] (e),
   (a)--[dotted] (d),
   (d)--[dotted] (b),
   (a)--[dotted](i),
   (e)--[photon, thick, rmomentum=\(-\ell\)] (j),
   (e)--[photon,thick,momentum'=\(q-\ell\)] (k),
   (j)--[dotted] (g),
   (j)--[dotted] (h)};
   \end{feynman}
   \end{tikzpicture}
   \qquad
   \begin{tikzpicture}
   \begin{feynman}
   \vertex[dot] at ($(0,0)$) (a){};
   \vertex[dot] at ($(0,2)$) (b){};
   \vertex at ($(-1,0)$) (c) {2};
   \vertex[dot] at ($(1,0)$) (d){};
   \vertex at ($(1,2)$) (e);
   \vertex at ($(2,0)$) (f);
   \vertex at ($(-1,2)$) (g) {1};
   \vertex at ($(2,2)$) (h);
   \vertex[isosceles triangle, draw, fill=black, rotate=45 ,scale=0.4 ] at ($(2,1)$) (i) {};
   \diagram*{
   (a)--[dotted] (c),
   (a)--[fermion, thick, edge label=\(\tilde{\omega}\)] (d),
   (d)--[dotted] (f),
   (a)--[photon, thick, rmomentum=\(-\ell\)] (b),
   (b)--[dotted] (g),
   (b)--[dotted] (e),
   (d)--[photon, thick, momentum=\(q-\ell\)] (i),
   (e)--[dotted] (h),};
   \end{feynman}
   \end{tikzpicture}
   \qquad
   \begin{tikzpicture}
   \begin{feynman}
   \vertex[dot] at ($(0,0)$) (a){};
   \vertex[dot] at ($(0,2)$) (b){};
   \vertex at ($(-1,0)$) (c) {2};
   \vertex at ($(1,0)$) (d);
   \vertex[dot] at ($(1,2)$) (e){};
   \vertex at ($(2,0)$) (f);
   \vertex at ($(-1,2)$) (g) {1};
   \vertex at ($(2,2)$) (h);
   \vertex[isosceles triangle, draw, fill=black, rotate=-45 ,scale=0.4 ] at ($(2,1)$) (i) {};
   \diagram*{
   (a)--[dotted] (c),
   (a)--[dotted] (d),
   (d)--[dotted] (f),
   (a)--[photon, thick, momentum=\(\ell\)] (b),
   (b)--[dotted] (g),
   (b)--[fermion, thick, edge label=\(\tilde{\omega}\)] (e),
   (e)--[photon, thick, momentum'=\(\ell-q\)] (i),
   (e)--[dotted] (h),};
   \end{feynman}
   \end{tikzpicture}}
   \scalemath{0.76}{
   \begin{tikzpicture}
   \begin{feynman}
   \vertex[dot] at ($(0,0)$) (a){};
   \vertex[dot] at ($(0,2)$) (b){};
   \vertex at ($(-1,0)$) (c) {2};
   \vertex at ($(1,0)$) (d){};
   \vertex at ($(1,2)$) (e);
   \vertex at ($(2,0)$) (f);
   \vertex at ($(-1,2)$) (g) {1};
   \vertex at ($(2,2)$) (h);
   \vertex[isosceles triangle, draw, fill=black, rotate=45 ,scale=0.4 ] at ($(1,1)$) (i) {};
   \diagram*{
   (a)--[dotted] (c),
   (a)--[dotted] (d),
   (d)--[dotted] (f),
   (a)--[photon, thick, rmomentum=\(-\ell\)] (b),
   (b)--[dotted] (g),
   (b)--[dotted] (e),
   (a)--[photon, thick, momentum'=\(q-\ell\)] (i),
   (e)--[dotted] (h),};
   \end{feynman}
   \end{tikzpicture}
   \qquad
   \begin{tikzpicture}
   \begin{feynman}
   \vertex[dot] at ($(0,0)$) (a){};
   \vertex[dot] at ($(0,2)$) (b){};
   \vertex at ($(-1,0)$) (c) {2};
   \vertex at ($(1,0)$) (d);
   \vertex at ($(1,2)$) (e){};
   \vertex at ($(2,0)$) (f);
   \vertex at ($(-1,2)$) (g) {1};
   \vertex at ($(2,2)$) (h);
   \vertex[isosceles triangle, draw, fill=black, rotate=-45 ,scale=0.4 ] at ($(1,1)$) (i) {};
   \diagram*{
   (a)--[dotted] (c),
   (a)--[dotted] (d),
   (d)--[dotted] (f),
   (a)--[photon, thick, momentum=\(\ell\)] (b),
   (b)--[dotted] (g),
   (b)--[dotted] (e),
   (b)--[photon, thick, momentum=\(\ell-q\)] (i),
   (e)--[dotted] (h),};
   \end{feynman}
   \end{tikzpicture}}
   \end{center}
\caption{Diagrams contributing to the sub-leading metric $h_{\mu\nu}^{(2)}$.
The top diagram carries $m_2^2$, and represents a 2PM contribution to the Kerr metric;
the five remaining diagrams carry $m_1m_2$, and represent the leading-order scattering waveform
of two spinning black holes first derived in Ref.~\cite{Jakobsen:2021lvp}.}
\label{waveformDiags}
\end{figure}

Our other main check is the on-shell condition $g_{\mu\nu}(x_1)\dot{x}_1^\mu\dot{x}_1^\nu=1$.
At 2PM order, expansion of the on-shell condition
in terms of the background fields~\eqref{backgroundExp} yields
\begin{align}\label{subleadingOS}
0&=2v_1\cdot\dot{z}_1^{(2)}+(\dot{z}_1^{(1)})^2+2h^{(1)}_{\mu \nu}(b^\mu_1+\tau_1 v^\mu_1)v_1^\mu\dot{z}_1^{(1)\nu}\\
&\quad+ h^{(2)}_{\mu\nu}(b^\mu_1+\tau_1 v^\mu_1)v_1^\mu v_1^\nu+\partial_\rho h^{(1)}_{\mu\nu}(b^\mu_1+\tau_1 v^\mu_1)v_1^\mu v_1^\nu z^{(1)\rho}_1\,.\nn
\end{align}
Besides the trajectory corrections $z_1^{(n)}$,
this check requires inclusion of the metric corrections $h_{\mu\nu}^{(n)}$.
The first correction $h_{\mu\nu}^{(1)}$ was provided in Eq.~\eqref{leadingMetric};
the sub-leading correction is represented by the six diagrams in Fig.~\ref{waveformDiags}.
Like the trajectory, these diagrams are organised into mass sectors carrying either $m_2^2$ ---
representing a sub-leading contribution to the Kerr metric ---
or $m_1m_2$, whose contributions represent the leading-order gravitational waveform from a scattering
of two spinning black holes~\cite{Jakobsen:2021lvp}.

The check on the sub-leading on-shell condition is
most easily performed in momentum space.
Inserting the Fourier transform on the trajectory~\eqref{trajFourier}
the condition~\eqref{subleadingOS} is rendered as
\begin{align}\label{OSmomentum}
   &0=\int_q e^{iq\cdot\tilde b}\dd(q\cdot v_2)\bigg\{2(-\i q\cdot v_1)v_1\cdot z_1^{(2)}(q)+h_{\mu\nu}^{(2)}(q)v_1^\mu v_1^\nu\nn\\
   &+\int_\ell\dd(\ell\cdot v_2)\Big[(-\i \ell\cdot v_1)(-\i(q\!-\!\ell)\cdot v_1)z_1^{(1)}(\ell)\cdot z_1^{(1)}(q\!-\!\ell)\nn\\
   &+\big(2(-\i(q\!-\!\ell)\cdot v_1)\delta^\nu_\rho+(-\i \ell_\rho)v_1^\nu \big)v_1^\mu h_{\mu\nu}^{(1)}(\ell)z_1^{(1)\rho}(q\!-\!\ell)\Big]\bigg\}.
\end{align}
Here we introduced the notation
\begin{equation}
  h^{(n)}_{\mu\nu}(b_1^\mu+\tau_1v_1^\mu)=
  \int_k e^{\i k\cdot \tilde{b}(\tau_1)}\deltabar(k\cdot v_2)h^{(n)}_{\mu\nu}(k)\,.
\end{equation}
Time derivatives on the trajectory are replaced by
factors of $-\i (q\cdot v_1)$ while derivatives $\partial_\mu$ on the waveform
are replaced by $-\i q^\mu$.
We have also used
\begin{equation}\label{convolutionThm}
\int_{q_1} e^{\i q_1\cdot\tilde b} g(q_1) \int_{q_2}e^{\i q_2\cdot\tilde b} h(q_2)
=\int_{q,\ell}e^{\i q\cdot\tilde b} g(\ell) h(q-\ell)\ ,
\end{equation}
which is derived by identifying $q^\mu=q^\mu_1+q^\mu_2$ and $\ell^\mu=q^\mu_1$.
We need only demonstrate vanishing of the terms in curly brackets of Eq.~\eqref{OSmomentum}
to verify the on-shell condition~\eqref{subleadingOS}.

% {\color{red}
% \begin{equation}
% \begin{split}
% &0=2(-\i q\cdot v_1)v_1\cdot z_1^{(2)}(q)\\
% & -\int_{-\infty}^\infty ds \big((s\cdot v_1)((q-s)\cdot v_1)z^\mu_{(1)}(s)z(q-s)_{\mu, (1)}\\
% &-\i2(s\cdot v_1)z(s)_{(1)}^\mu h(x(q-s))_{\mu \nu,(1)}v_1^\nu\\
% &+\i (q-s)_\rho h(x(q-s))_{\mu\nu,(1)}z^{\rho}_{(1)}(s)v_1^\mu v_1^\nu\big)\\
% &+h(x(q))_{\mu\nu, (2)}v_1^\mu v_1^\nu\, .
% \end{split}
% \end{equation}}

In order to verify Eq.~\eqref{OSmomentum},
we must perform the integral on $\ell^\mu$ present for certain terms.
Quite conveniently, the integrals involved all belong to the one-loop probe family $I^{1-\mathrm{loop}}_{\nu_1,\nu_2,\nu_3}$~\eqref{probeFamily}
already used when evaluating the 2PM trajectory.
The only new subtlety is that these one-loop integrals on $\ell^\mu$
can have worldline propagators of the kind $1/(\ell\cdot v_1+\i 0)$
and $1/((q-\ell)\cdot v_1+\i 0)$.
In the latter case we can simply relabel $\ell\to q-\ell$ to bring our integrals into the form of Eq.~\eqref{probeFamily};
however, if both worldline propagators appear together then we have an over complete set.
We therefore apply the partial fraction identity:
\begin{align}
&\frac{1}{(\ell\cdot v_1+\i0)((q-\ell)\cdot v_1+\i0)}=\\
&\qquad\frac{1}{(q\cdot v_1+\i0)}\bigg(\frac{1}{(q-\ell)\cdot v_1+\i0}+\frac{1}{\ell\cdot v_1+\i0}\bigg)\,,\nn
\end{align}
and $(q\cdot v_1+\i0)^{-1}$ can be pulled outside the integral.
The resulting terms can then be relabelled as above.
% In the second case, the integral family is related to $I^{1-\mathrm{loop}}_{\nu_1,\nu_2,\nu_3}$ by a shift $s\to s+q$ and a variable change $s\to -s$. This variable change will not affect terms in the result which are even in $q$, but only terms which are odd in $q$ i.e. terms containing odd powers of $q\cdot v_1$. Note that we do not take the asymptotic limit in this calculation. It is valid across all time.
Given these ingredients, we can confirm that the on-shell condition~\eqref{subleadingOS} holds as
expected up to linear order in spin.

The involvement of the full gravitational scattering waveform makes this highly non-trivial.
Similarly, we can verify local preservation of the spin-supplementary condition (SSC)
$\dot{x}_{i,\mu}S_i^{\mu\nu}=0$,
which follows from
\begin{align}
   \alpha_i\cdot \pi_i=0\, ,
\end{align}
with the covariant momentum
 $\pi^\mu_i=m_i\dot{x}^\mu_i+\mathcal{O}(R^2)$ \cite{Haddad:2024ebn}.
% At the spin order we are currently dealing with the second term can be safely ignored,
% and
The analog of Eq.~\eqref{OSmomentum} is
\begin{align}
&0=\int_q e^{\i q\cdot \tilde{b}}\dd(q\cdot v_2)
\bigg\{(-\i q\cdot v_1)\alpha_{1,\,-\infty}\cdot z_1^{(2)}(q)+\alpha_1^{(2)}(q)\cdot v_1\nn\\
&\quad+\int_\ell\dd(\ell\cdot v_2)(-\i (q-\ell)\cdot v_1)\alpha_1^{(1)}(\ell)z_1^{(1)}(q-\ell)\bigg\}\,.
\end{align}
This involves the Fourier transforms of $\alpha_1$, $\bar{\alpha}_1$ and $z_1$.
By similar calculations we have moreover confirmed that
\begin{align}\label{charges}
\begin{aligned}
(\alpha_1)^2=\text{const}, \quad (\bar{\alpha}_1)^2=\text{const}, \quad \bar{\alpha}_1\cdot \alpha_1=\text{const}.
\end{aligned}
\end{align}
These quantities are related to the spin length by
\begin{equation}
-\frac{1}{2}S_1^{\mu\nu}S_{1,\mu\nu}=(\bar{\alpha}_1\cdot \alpha_1)^2-\bar{\alpha}_1^2\alpha_1^2\,,
\end{equation}
and so the spin length is itself also conserved.

\section{Radiated Angular Momentum}\label{sec:jRad}

Using our new results for the full scattering trajectories,
let us finally compute the overall change in angular momentum $\Delta J^\mu$ ---
and thus the total radiated angular momentum $J^\mu_{\rm rad}=-\Delta J^\mu$.
The angular momentum vector $J^\mu$, defined with respect to the COM frame, is
\begin{equation}\label{defJ}
J^\mu=\sfrac{1}{2}{\epsilon^\mu}_{\nu \rho \sigma}J^{\nu \rho}\hat{P}^\sigma,
\end{equation}
where $\hat{P}^\mu=\frac{1}{E}(m_1 v_1^\mu+m_2 v_2^\mu)$
is the unit vector pointing in the direction of the total momentum,
$E=M\sqrt{1+2\nu(\gamma-1)}$ is the total incoming energy and $M=m_1+m_2$, $\nu=m_1m_2/M^2$.
At the perturbative PM orders we are dealing with in this paper,
and given that momentum loss due to gravitational wave emission begins only at 3PM order \cite{Damour:1981bh,Damour:2020tta},
we may safely assume that the total energy $E$ and momentum $P^\mu$
(and therefore $\hat{P}^\mu$) are conserved.
The total angular momentum tensor is given by a sum of orbital and spin components:
\begin{align}
   J^{\mu\nu}&=L^{\mu\nu}+S_1^{\mu\nu}+S_2^{\mu\nu}\,,\\
   L^{\mu\nu}&=2x_1^{[\mu}p_1^{\nu]}+2x_2^{[\mu}p_2^{\nu]}\,.
\end{align}
The two parts of the orbital angular momentum,
labelled respectively by the subscripts ``$1$" and ``$2$",
are parametrised by separate time parameters $\tau_i$.
Inserting into Eq.~\eqref{defJ} we learn that
% \begin{subequations}
% \begin{align}\label{eq:jVec}
%    J^\mu=L^\mu+\sum_i\left(\dot{x}_i\cdot \hat P\,S_i^\mu-S_i\cdot \hat P\,\dot{x}_i^\mu\right)\,,
% \end{align}
% \end{subequations}
\begin{subequations}\label{eq:jVec}
   \begin{align}
      J^\mu&=L^\mu+\sum_{i=1}^2{{\Lambda_i}^\mu}_\nu S_i^\nu\,,\\
      {{\Lambda_i}^\mu}_\nu&=
      \dot{x}_i\cdot \hat P\,\delta^\mu_\nu-\dot{x}_i^\mu\hat P_\nu\,,
   \end{align}
\end{subequations}
where the angular momentum vector $L^\mu$ is given by
\begin{align}\label{LVec}
   L^\mu=\sfrac12{\eps^\mu}_{\nu\rho\sigma}L^{\nu\rho}\hat{P}^\sigma
   =\sfrac{1}{E} {\eps^\mu}_{\nu\rho\sigma}(x_1-x_2)^\nu p_1^\rho p_2^\sigma\,.
\end{align}
Notice that $J^\mu\neq L^\mu+S_1^\mu+S_2^\mu$,
because the spin vectors $S_i^\mu$ are defined with respect to their
respective rest frames $\dot{x}_i^\mu$~\eqref{SVec}.
The boosts ${{\Lambda_i}^\mu}_\nu$ are therefore needed to
bring them into the center-of-mass frame where $L^\mu$ is also defined.

The asymptotic change $\Delta J^\mu$ is given by
\begin{equation}
\label{asymptotic}
   % \Delta J^\mu=J^\mu(t_{\mathrm{COM}}=\infty)-J^\mu(t_{\mathrm{COM}}=-\infty),
   \Delta J^\mu=J^\mu(\tau_1, \tau_2\to+\infty)-J^\mu(\tau_1, \tau_2\to-\infty)\,,
\end{equation}
and using Eq.~\eqref{eq:jVec} we have
\begin{align}\label{deltaJ}
   &\Delta J^\mu=\Delta L^\mu+\sum_i\frac{1}{m_i}\left(\Delta p_i\cdot \hat P\,S_i^\mu-S_i\cdot \hat P\,\Delta p_i^\mu\right)\\
   &+\sum_i\frac{1}{m_i}\left((p_i+\Delta p_i)\cdot \hat P\,\Delta S_i^\mu-\Delta S_i\cdot \hat P\,(p_i^\mu+\Delta p_i^\mu)\right).\nn
\end{align}
This involves the change in the orbital angular momentum vector $\Delta L^\mu$.
As with the trajectory, we compute $\Delta J^\mu$ at leading (1PM) and sub-leading (2PM) order.
At leading order, this merely confirms that no angular momentum is radiated,
thus providing another check on the trajectory.
\subsection{Leading order}

At leading-PM order,
we can write down the angular momentum as a function of the proper times of both massive bodies $\tau_{1,2}$.
In the non-spinning case, inserting the time-dependent trajectory into Eq.~\eqref{LVec}
we have
\begin{align}
     &L_{S^0}^{(1)\mu}(\tau_{1},\tau_{2})=\\
     &\quad\frac{ m_1 m_2 |b|  \hat{l}^\mu [\gamma  (3-2 \gamma ^2) m_1+(3-4 \gamma ^2) \
m_2]}{2 \gamma v 
D(\tau_1)E}+ (1 \leftrightarrow 2)\, .\nn
  \end{align}
It is clear that the difference
$\Delta L^\mu=L^\mu(+\infty,+\infty)-L^\mu(-\infty,-\infty)$ vanishes,
as  $D(\tau_i)=D(-\tau_i)$.
Thus, as $\Delta J^\mu=\Delta L^\mu$ there is no dissipation and angular momentum is conserved.
Working up to quadratic order in spin,
we require both the orbital and spin contributions to the total angular momentum $J^\mu$.
The orbital component is presented here up to linear order in spin:
\begin{widetext}
   \begin{align}
 &L^{(1)\mu}_{S^1}(\tau_1,\tau_2)=
 \frac{ m_1 m_2}{\hat{D}(\tau_1)^3 |b|E} \bigg[\frac{ m_ 2 \hat{b}^\mu(\gamma ^2 v a_ 1\cdot \hat{b}-a_ 1\cdot v_ 2 
(\hat{D}(\tau_1)^3+\gamma ^3 \hat{\tau}_1^3 v^3)-2 \gamma ^4 \hat{\tau}_1^2 v^3 a_ 2\cdot \hat{b}
+2 \gamma  a_ 2\cdot v_ 1 (\hat{D}(\tau_1)^3+\gamma ^3 \hat{\tau}_1^3 v^3))}{\gamma 
 v}\nn\\
 &+\frac{(\gamma  m_1+m_2) w_1^\mu(\gamma ^2 v a_1\cdot 
\hat{b} (2 \hat{D}(\tau_1)^3+2 \gamma ^3 \hat{\tau}_1^3 
v^3+3 \gamma  \hat{\tau}_1 v)-a_1\cdot v_2+2 \gamma  (\gamma  v 
a_2\cdot \hat{b} (\hat{D}(\tau_1)^3+\gamma ^3 
\hat{\tau}_1^3 v^3+2 \gamma  \hat{\tau}_1 v)+a_2\cdot v_1))}{2 
}\nn\\
&-\frac{ m_ 1v_2^\mu(\gamma ^2 v a_ 1\cdot \hat{b} (2 \hat{D}(\tau_1)^3
+2 \gamma ^3 \hat{\tau}_1^3 v^3+3 \gamma  \hat{\tau}_1 v)-a_ 1\cdot 
v_ 2+2 \gamma  (\gamma  v a_ 2\cdot \hat{b} (\hat{D}(\tau_1)^3
+\gamma ^3 \hat{\tau}_1^3 v^3+2 \gamma  \hat{\tau}_1 v)+a_ 2\cdot 
v_ 1))}{2}\nn\\
&+\frac{\hat{\ell}^\mu(a_ 1\cdot \hat{\ell} (m_ 1-\gamma  m_ 2)-2 \gamma  a_2\cdot \hat{\ell} 
(\gamma  m_ 1 v^2+2 m_ 2 (\gamma ^2 \hat{\tau}_1^2 v^2+1)))}{2 
}\bigg]+(1 \leftrightarrow 2)\,,
 \end{align}
where $\hat{D}(\tau)$ was introduced in Eq.~\eqref{Dtau}.
Meanwhile, the spin contribution is given by
\begin{align}
&\sum_i{{\Lambda_i}^\mu}_\nu S_i^{(1)\nu}=\\
&\frac{ m_1 m_2}{|b|E}\bigg[
\hat{b}^\mu\Big[\frac{2 (m_1+\gamma  m_2) (a_1\cdot v_2 (2 
\gamma ^2 \hat{D}(\tau_1) \hat{\tau}_1^2 
v^2+\hat{D}(\tau_1)+2 \gamma ^3 \hat{\tau}_1^3 v^3+2 \gamma  
\hat{\tau}_1 v)-v a_1\cdot \hat{b} (\gamma  \hat{D}(\tau_1) 
\hat{\tau}_1 v+\gamma ^2 \hat{\tau}_1^2 
v^2+1))}{\hat{D}(\tau_1)^2 v S(\tau_1)}\nn\\
&-\frac{a_1\cdot \hat{b} ((2 \gamma ^2-1) m_1+2 
\gamma  m_2) (\hat{D}(\tau_1)+\gamma  \hat{\tau}_1 
v)}{\gamma  \hat{D}(\tau_1) \hat{\tau}_1 v+\gamma ^2 
\hat{\tau}_1^2 v^2+1}-\frac{(2 \gamma ^2-1) m_2 a_1\cdot v_2 
(\hat{D}(\tau_1)+\gamma  \hat{\tau}_1 v)}{\gamma v 
\hat{D}(\tau_1)}\Big]\nn\\
&-\frac{ (\gamma  m_1\!+\!m_2)w_1^\mu (2 \gamma ^2 v a_1\cdot \hat{b} (4 
\gamma ^2 \hat{D}(\tau_1) \hat{\tau}_1^2 v^2\!+\!\hat{D}(\tau_1)\!+\!4 
\gamma ^3 \hat{\tau}_1^3 v^3\!+\!3 \gamma  \hat{\tau}_1 v)+(4 \gamma ^2\!-\!1) 
a_1\cdot v_2 (2 \gamma  \hat{D}(\tau_1) \hat{\tau}_1 v+2 \gamma ^2 
\hat{\tau}_1^2 v^2+1))}{S(\tau_1)
(\gamma  \hat{D}(\tau_1) \hat{\tau}_1 v+\gamma ^2 \hat{\tau}_1^2 
v^2+1)}\nn\\
&+\frac{ m_1v_2^\mu(2 \gamma ^2 v a_1\cdot \hat{b}\, (4 \gamma ^2 
\hat{D}(\tau_1) \hat{\tau}_1^2 v^2+\hat{D}(\tau_1)+4 
\gamma ^3 \hat{\tau}_1^3 v^3+3 \gamma  \hat{\tau}_1 v)+(4 \gamma 
^2-1) a_1\cdot v_2 (2 \gamma  \hat{D}(\tau_1) \hat{\tau}_1 
v+2 \gamma ^2 \hat{\tau}_1^2 v^2+1))}{
S(\tau_1) (\gamma  
\hat{D}(\tau_1) \hat{\tau}_1 v+\gamma ^2 \hat{\tau}_1^2 
v^2+1)}\nn\\
&+\frac{ a_1\cdot \hat{\ell}\, \hat{\ell}^\mu}{ (\gamma v\hat{\tau}_1
S(\tau_1)+|b|)} \Big[-\frac{2 (m_1+\gamma  m_2)
(\gamma  \hat{D}(\tau_1) \hat{\tau}_1 v+\gamma ^2 
\hat{\tau}_1^2 v^2+1)^2}{\hat{D}(\tau_1)^2 
S(\tau_1)}- ((2 \gamma 
^2-1) m_1+2 \gamma  m_2) S(\tau_1)\Big]\bigg]+(1 \leftrightarrow 2)\, .\nn
   % &\sum_i{{\Lambda_i}^\mu}_\nu S_i^\nu=
   % \frac{G m_1 m_2}{ 
   % \hat{D}(\tau_1) |b|\gamma ^2 v^2 S(\tau_1)} \Big[S(\tau_1) 
   % (\gamma ^2 v^2 \hat{l}^\mu\, a_1\cdot \hat{l} \big[2 \gamma ^2 m_1+m_1+4 \gamma  
   % m_2\big]\\
   % &-(v_2^\mu (m_1+\gamma  m_2)-v_1^\mu (\gamma  m_1+m_2)) \big[2 
   % \gamma ^3 \hat{\tau_1} v^2 a_1\cdot \hat{b}+(2 \gamma ^2 
   % (v^2+1)+1) a_1\cdot v_2\big])\\
   % &-\gamma  v \hat{b}^\mu \big[\gamma  v 
   % a_1\cdot \hat{b} (2 \gamma ^2 m_1+m_1+4 \gamma  m_2) 
   % S(\tau_1)+a_1\cdot v_2 (m_2 (2 
   % \gamma ^2+2 \gamma ^3 \hat{D}(\tau_1) \hat{\tau_1} v^3+2 \gamma 
   % ^4 \hat{\tau_1}^2 v^4-1)-2 \gamma ^2 m_1 \hat{\tau_1} v 
   % S(\tau_1))\big]\Big]\\
   % &+ (1 \leftrightarrow 2)\, .
\end{align}
\end{widetext}
Full results up to quadratic order in spin are provided in the \zenodo submission.
Summing up these two contributions to obtain $J^\mu(\tau_1,\tau_2)$~\eqref{eq:jVec},
we have confirmed that $\Delta J^\mu=0$ at this perturbative order,
and thus there is no radiated angular momentum.

\subsection{Subleading order}

Finally, in this section we calculate $\Delta J^\mu$ up to $\cO(G^2)$ precision,
i.e.~a one-loop result up to linear order in spin.
As our result for the sub-leading trajectory~\eqref{subleadingTr} is given only in momentum space,
we here take a slightly different approach than used at leading order 1PM.
The change in the orbital angular momentum vector $L^\mu$ is given by
\begin{align}\label{deltaL}
   \Delta L^\mu=\sfrac12{\eps^\mu}_{\nu\rho\sigma}\Delta L^{\nu\rho}\hat{P}^\sigma\,.
\end{align}
Meanwhile, the 2PM expression for the angular momentum tensor $L^{\mu\nu}$ decomposes as follows:
\begin{align}\label{Lsplit}
  L^{(2)\mu\nu}(\tau_1,\tau_2)
  &=2\sum_{i=1}^2\big(x^{(2)[\mu}_i(\tau_i) p^{(0)\nu]}_i\!+\!x^{(1)[\mu}_i(\tau_i) p^{(1)\nu]}_i(\tau_i)\nn\\
  &\qquad\qquad+x^{(0)[\mu}_i(\tau_i) p^{(2)\nu]}_i(\tau_i)\big)\, .
\end{align}
We evaluate these three contributions separately,
ultimately obtaining results for the radiated angular momentum in agreement with the literature~\cite{Bini:2024hme,Jakobsen:2022fcj}.

To evaluate the first contribution, it suffices to plug in our momentum space result
and take the asymptotic limit.
The procedure is analogous to Eq.~\eqref{limitfordp}:
% \begin{equation}
%    \begin{split}
%    & x_{\text{2PM}}^{[\mu}\, p_{\text{0PM}}^{\nu]}=\\
%    &\qquad-m_1\left(\int_q e^{\i q\cdot b} \delta(q\cdot v_1)z(q)^{[\mu}_{1,\, 2\mathrm{PM}}\right) v_1^{\nu]}\ ,
%    \end{split}
% \end{equation}
\begin{align}
   &\Delta\big(x_1^{(2)[\mu} p_1^{(0)\nu]}\big)\\
   &=m_1\int_{-\infty}^\infty\d\tau_1\frac{\d}{\d\tau_1}\bigg(\int_q e^{\i q\cdot\tilde{b}(\tau_1)}\dd(q\cdot v_2)z_1^{(2)[\mu}(q)v_1^{\nu]}\bigg)\nn\\
   &=m_1\int_q e^{\i q\cdot b}\dd(q\cdot v_1)\dd(q\cdot v_2)(-iq\cdot v_1)z_1^{(2)[\mu}(q)v_1^{\nu]}\,.\nn
\end{align}
For concreteness, we have specialized to the trajectory of BH1.
Tackling the second term in Eq.~\eqref{Lsplit} involves computing products of Fourier transforms,
so we again make use of convolutions~\eqref{convolutionThm}:
\begin{align}
   \label{Jconvolution}
   &\Delta\big(x_1^{(1)[\mu} p_1^{(1)\nu]}\big)\\
   &=\Delta\bigg(\int_{q,\ell}e^{\i q\cdot\tilde{b}(\tau_1)}
   \dd(q\cdot v_2)\dd(\ell\cdot v_2)z_1^{(1)[\mu}(\ell)p_1^{(1)\nu]}(q-\ell)\bigg)\nn\\
   &=\int_q e^{\i q\cdot b}\dd(q\cdot v_1)\dd(q\cdot v_2)(-iq\cdot v_1)\nn\\
   &\qquad\qquad\times\int_\ell\dd(\ell\cdot v_2) z_1^{(1)[\mu}(\ell)p_1^{(1)\nu]}(q-\ell)\,.\nn
\end{align}
% \begin{align}
% \int_q e^{\i q\cdot \tilde{b}_i} \int_{\hat{q}}e^{\i \hat{q}\cdot \tilde{b}_i}\,x_i^{[\mu}(q) \,p_i^{\nu]}(\hat{q})=\\
%  \int_q \int_s e^{\i q\cdot \tilde{b}_i }\, x^{[\mu}_i(s)\, p^{\nu]}_i(q-s)\ .
% \end{align}
The resulting one-loop integral family, again, is the same encountered earlier.
While it is also feasible to obtain this contribution in position space,
we perform the calculation in momentum space in order to maintain the link to higher perturbative orders.
As for the third term in Eq.~\eqref{Lsplit},
using the zeroth-order trajectory $x_1^{(0)\mu}=b_1^\mu+\tau_1v_1^\mu$
we have
\begin{equation}
   \begin{split}
   &\Delta(x_1^{(0)[\mu}\, p_1^{(2)\nu]})=
   b_1^{[\mu} \Delta p_1^{(2)\nu]}+v_{1}^{[\mu}\Delta(\tau_1 p_1^{(2)\nu]})\ .
   \end{split}
\end{equation}
The second term is evaluated using an integration-by-parts identity. We decompose it as
\begin{align}\label{deltaprime}
   &v_{1}^{[\mu}\Delta(\tau_1 p_1^{(2)\nu]})=
   v_{1}^{[\mu}\!\!\!\int_{-\infty}^{\infty}\!\!\!\!\d\tau \frac{\d}{\d\tau}\bigg(\!\tau\!\int_q e^{\i q\cdot \tilde{b}(\tau)}\dd(q\!\cdot\!v_2)p^{(2)\nu]}(q)\!\bigg)\nn\\
   % &=v_{1}^{[\mu}\bigg(\int_q\dd(q\cdot v_1)\dd(q\cdot v_2)e^{\i q\cdot b}p^{(2)\nu]}(q)\\
   % &\qquad+\int_q \dd^\prime(q\cdot v_1)(q\cdot v_1)\dd(q\cdot v_2)e^{\i q\cdot b}p^{(2)\nu]}(q)\bigg)\ .\\
   &=v_{1}^{[\mu}\int_q e^{\i q\cdot b}\big(\dd(q\!\cdot\! v_1)+(q\!\cdot\! v_1)\dd^\prime(q\!\cdot\! v_1)\big)\dd(q\!\cdot\! v_2)p^{(2)\nu]}(q)\nn\\
   &=-v_1^{[\mu}\int_q e^{\i q\cdot b}\dd(q\!\cdot\!v_1)\dd(q\!\cdot\!v_2)
   (q\!\cdot\!v_1)\frac{\mathrm{d}}{\mathrm{d}(q\!\cdot\!v_1)}\big[p(q)^{(2)\nu]}\big]\,.
\end{align}
% The last term in the above we integrate by parts:
% \begin{align}
% \begin{aligned}
% &v_1^{[\mu}\int_q  \dd^\prime(q\cdot v_1)(q\cdot v_1)\dd(q\cdot v_2) e^{\i q\cdot b} p(q)^{(2)\nu]} \\
% =&-v_1^{[\mu}\bigg(\int_q \dd(q\cdot v_1) (q\cdot v_1) \dd(q\cdot v_2)e^{\i q\cdot b}\frac{\mathrm{d}}{\mathrm{d}(q\cdot v_1)}\big[p(q)^{(2)\nu]}\big]\\
% &-\int_q \dd(q\cdot v_1)\dd(q\cdot v_2)e^{\i q\cdot b}p(q)^{(2)\nu]}\bigg)\, ,
% \end{aligned}&&
% \end{align}
% Here the last term cancels against the first term in \eqref{deltaprime}.
The remaining contribution is a Fourier integral over
$v_{1}\cdot \partial_{q}p(q)^{(2)\nu}$ by virtue of the chain rule.

Adding all contributions to the angular momentum tensor~\eqref{Lsplit} together,
and combining to obtain the change in angular momentum vector $\Delta L^\mu$~\eqref{deltaL},
we find
\begin{widetext}
\begin{equation}\label{radJ2PM}
\begin{split}
&\Delta J^{(2)\mu}_{S^0}=-\frac{m_1 m_2\, \, 4(2\gamma^2-1)}{3|b|(\gamma v)^{3}} \big[(8-5\gamma^2)\gamma v+3\gamma (2\gamma^2-3)\arccosh(\gamma)\big] \hat{l}^\mu\\
&+\frac{8 i \pi\, m_1 m_2\,  \gamma(3-8\gamma^2+4\gamma^4)}{|b| (\gamma v)^{3}}\bigg[\int_q \deltabar (q\cdot v_1)\deltabar(q\cdot v_2) (b\cdot q)\Big(e^{i q\cdot b}\Big(\frac{1}{2\epsilon}-\frac{\i\pi}{2}-\log[(\hat{q}\cdot v_1)_N^{-1}]\Big)+(1 \leftrightarrow 2)\Big)\bigg] \hat{l}^\mu \,,
%\mathcal{O}(S^1) \ .
\end{split}
\end{equation}
\end{widetext}
where in this case $\Delta J^\mu=\Delta L^\mu$ due to the absence of spin.
The entire second line cancels by symmetry:
as the result can only depend on $|b|$ and $\gamma=v_1\cdot v_2$,
one may freely exchange particle labels: $v^\mu_1\leftrightarrow v^\mu_2$
and $b^\mu_1\leftrightarrow b^\mu_2$, implying $b^\mu\to-b^\mu$.
The final result, given in the top line of Eq.~\eqref{radJ2PM},
gains contributions only from the $m_1m_2$ mass sector of the 2PM trajectory.
This aligns with our expectations, as it is this mass sector which contains diagrams where gravitons can go on-shell.

% \begin{equation}
% \begin{split}
% &\Delta J^{(2)\mu}_{S^0}=\frac{m_1 m_2\, G^2\, 4(2\gamma^2-1)}{3|b|^2(\gamma^2-1)^{5/2}} \big[-(8-13\gamma^2+5\gamma^4)\\
% &\qquad+3\gamma \sqrt{\gamma^2-1}(2\gamma^2-3)\arccosh(\gamma)\big]|b| P_{\mathrm{COM}}\hat{l}^\mu\ .
% \end{split}
% \end{equation}

The only substantial difference in the spinning case,
besides the higher tensor rank of the integrals involved (loop integrals and Fourier transforms),
is that we must also include contributions from the change in spin vectors~\eqref{eq:jVec}.
Up to linear order in spin,
\begin{align}
\begin{aligned}
\Delta \bigg(\sum_{i=1}^2{{\Lambda_i}^\mu}_\nu S_i^\nu\big)=\Delta\bigg(\sum_{i=1}^2 \dot{x}_{i, S^0}\cdot \hat{P}\, S_{i, S^1}^\mu
-\dot{x}_i^\mu\hat P\cdot S_{i, S^1}
\bigg).
\end{aligned}
\end{align}
Decomposing this into the contributing PM orders of the individual terms,
we find that at 2PM
\begin{align}
&\Delta\bigg(\sum_{i=1}^2{{\Lambda_i}^\mu}_\nu S_i^\nu\bigg)\\
&=\sum_{i=1}^2\bigg(\hat{P}\cdot \Delta p^{(2)}_i a_i^{\mu}+
(\hat{P}\cdot v_i)\Delta S_i^{(2) \mu}
+(\hat{P}\cdot \Delta p_i)\Delta a_i^{(1) \mu}\nn\\
&\quad-
\hat{P}\cdot \Delta S_i^{(2)}v_i^\mu-
(\hat{P}\cdot a_i)\Delta p_i^{(2) \mu}
-(\hat{P}\cdot \Delta a_i^{(1)})\Delta p_i^{(1) \mu}\bigg).\nn
\end{align}
Taking this together with the change in the orbital angular momentum vector,
we learn that overall:
\begin{align}
 &\Delta J^{(2)\mu}_{S^1}= -\frac{8 \gamma  m_1 m_2 }{3 (\gamma v)^2 |b|^2} \big[ (8-5 \gamma ^2)\gamma v\\
&+\gamma (6 \gamma ^2-9) \arccosh(\gamma )) (\hat{l}^\mu (a_{12}\cdot \hat{l})-b^\mu (a_{12}\cdot b)\big]\,.\nn
\end{align}
This result has components in the $b^\mu$ and $\hat{l}^\mu$ directions,
and agrees with the literature~\cite{Jakobsen:2021lvp,Jakobsen:2023hig}.
Divergent integrals of the kind in Eq.~\eqref{radJ2PM}
could be shown to cancel by using the symmetry of these terms under
the exchange of the black hole labels ``1" and ``2".

Finally, we specialize to aligned spins using
\begin{align}
a^\mu_{1}&=s_1 \hat{l}^\mu\,, &  a^\mu_2&=s_2 \hat{l}^\mu\,,
\end{align}
aligning the spin vectors with the orbital angular momentum of the system,
so $a_i\cdot v_j=0$ and $a_i\cdot b=0$.
The answer then simplifies to
\begin{equation}
\begin{split}
\Delta J^{(2)\mu}_{S^1}&=\frac{8 \gamma m_1 m_2 (s_1+s_2)}{3(\gamma v)^2|b|^2 }\big[(8-5\gamma^2)\gamma v\\
&\qquad+\gamma(6\gamma^2-9)\arccosh(\gamma)\big]\hat{l}^\mu\ .
\end{split}
\end{equation}
As in the non-spinning case, $\Delta J^\mu=\Delta L^\mu$ and the result points in the direction of $\hat{l}^\mu$.
There is no contribution due to the spin kick $\Delta S_i^\mu$ as this observable vanishes.

\section{Conclusions}

In this work, we have computed the trajectories of massive bodies involved
in a classical gravitational two-body scattering in the presence of spin,
using methods of perturbative quantum field theory.
We established the 1PM trajectories with spin effects
in both momentum and coordinate space, as well as the 2PM corrections in
momentum space including linear-in-spin contributions.
The underlying Feynman integrals contain an additional scale compared
to asymptotic observables: alongside the momentum transfer 
$q^\mu$ (dual to the impact parameter $b^\mu$)
we encountered the quantity $q\cdot v_{i}$ (dual to the time parameter $\tau_i$)
for the trajectory of black hole $i$. 
At 2PM order the Fourier transform to coordinate space remains challenging, mirroring
similar difficulties in the NLO waveform
\cite{Brandhuber:2023hhy,DeAngelis:2023lvf,Herderschee:2023fxh,
Caron-Huot:2023vxl,Bohnenblust:2023qmy,Bini:2024rsy,Elkhidir:2023dco,Bohnenblust:2025gir}.
Some of the ideas presented in 
Ref.~\cite{Brunello:2024ibk},
namely performing integration-by-parts (IBP) reductions in the presence of exponential Fourier factors, might aid future computations.
\newline
\newline
Our work demonstrates that calculating scattering trajectories
provides a novel pathway towards determining a crucial 
asymptotic quantity in the two-body scattering problem:
the radiated angular momentum.
Given the more cumbersome nature of the gravitational waveform,
it is advantageous to have a direct computational pathway to $J_{\rm rad}^\mu$ that
bypasses the waveform at higher perturbative orders.
% While previous approaches have relied on computing the full waveform
% \cite{Damour:2020tta,Jakobsen:2021smu,Jakobsen:2021lvp},
% the eikonal operator \cite{Heissenberg:2023uvo,Heissenberg:2024umh,Heissenberg:2025ocy}
% or linear response
% \cite{Bini:2012ji,Damour:2020tta,Bini:2021gat,Jakobsen:2022zsx,Jakobsen:2023hig},
Together with the radiated energy-momentum $P_{\rm rad}^\mu$,
$J_{\rm rad}^\mu$ will play a crucial role in the development of future gravitational
waveform models (including effective-one-body)
for the two-body problem informed by PM results~\cite{Antonelli:2019ytb,Khalil:2022ylj,Buonanno:2024vkx,Buonanno:2024byg,
Dlapa:2024cje,Damour:2025uka}.
% The ability to construct the radiated angular momentum alongside
% established efficient QFT based paths to the radiated
% momentum is of paramount importance for developing waveform models that
% incorporate fluxes—models essential for gravitational wave data analysis.

\section*{Acknowledgements}

We thank Riccardo Gonzo, Carlo Heissenberg, Gustav Uhre Jakobsen and Canxin Shi for useful discussions.
We also give special thanks to Benjamin Sauer and Johann Usovitsch
for their insights on multi-scale loop integrals. We thank the referee for 
pointing out the relation of our integrals to the waveform ones.
This work was funded by the Deutsche Forschungsgemeinschaft
(DFG, German Research Foundation)
Projektnummer 417533893/GRK2575 ``Rethinking Quantum Field Theory'' (GM, JP, KS), 
by The Royal Society under grant URF\textbackslash R1\textbackslash 231578
``Gravitational Waves from Worldline Quantum Field Theory'' (GM)
and  by the European Union through the 
European Research Council under grant ERC Advanced Grant 101097219 (GraWFTy) (JP).
Views and opinions expressed are however those of the authors only and do not necessarily reflect those of the European Union or European Research Council Executive Agency. Neither the European Union nor the granting authority can be held responsible for them.

\appendix

\begin{widetext}

\section{Leading-order trajectory at quadratic order in spin}

Here we present the leading-PM trajectory at quadratic order in spin.
We write the result in terms of its coefficients with respect to an orthogonal
basis spanned by $w_1^\mu$, $v_2^\mu$, $\hat{b}^\mu$ and $\hat{l}^\mu$. The coefficients are sorted by dot products of the spin vectors of both black holes.
\begin{equation}
   z_1^{(1)\mu}(\tau_1)_{S^2}=F_1 w_1^\mu+F_2 v_2^\mu+F_3 \hat{b}^\mu+F_4 \hat{l}^\mu\ .
\end{equation}
The coefficients are:
\begin{flalign}
   \begin{aligned}
      &F_1= \frac{ m_2}{\hat{D}(\tau_1)^3 |b|^2} \bigg[\frac{1}{2} a_1^2 \hat{\tau}_1 [3 
\gamma ^2+(2 \gamma ^4-3 \gamma ^2+1) \hat{\tau}_1^2-2]+a_1\cdot 
\hat{b} \Big[\gamma ^2 \hat{\tau}_1 (2 (\gamma v)^2 \hat{\tau}_1^2+3) 
a_2\cdot \hat{b}+\frac{a_2\cdot v_1}{v^2}\Big]\\
&+a_1\cdot v_2 
\Big[-\frac{a_1\cdot \hat{b}+a_2\cdot 
\hat{b}}{\gamma  v^2}-\gamma  \hat{\tau}_1 (\gamma ^2 
\hat{\tau}_1^2+1) a_2\cdot v_1\Big]+\gamma ^2 \hat{\tau}_1  a_1\cdot a_2[(\gamma v)^2 \hat{\tau}_1^2+1]\\
&+\frac{(a_1\cdot \hat{b})^2(2 \gamma ^4-3 \gamma 
^2+1) \hat{\tau}_1  (2 \gamma ^2 
\hat{\tau}_1^2 v^2+3)}{2 \gamma ^2 v^2}+\frac{(a_1\cdot v_2)^2}{2} \hat{\tau}_1 
[(2 \gamma ^2-1) \hat{\tau}_1^2+3]+\frac{1}{2} 
a_2^2 \hat{\tau}_1 [-2 \gamma ^2+(\gamma v)^2 
\hat{\tau}_1^2+3]\\
&+\frac{(a_2\cdot \hat{b})^2}{2} \hat{\tau}_1 [2 (\gamma v)^2 
\hat{\tau}_1^2+3]
+\frac{(2 \gamma ^2-1) 
a_2\cdot \hat{b}\, a_2\cdot v_1}{\gamma ^2 v^2}+\frac{1}{2} \hat{
\tau} (\hat{\tau}_1^2-4) (a_2\cdot 
v_1)^2
-\frac{a_1\cdot \hat{b}\, a_2\cdot v_1 (\gamma ^2 \hat{\tau}_1^2 
v^2+1)\hat{D}(\tau_1)}{v^2}\\
&+a_1\cdot v_2 \Big[\frac{(a_1\cdot \hat{b}+a_2\cdot \hat{b}) (\gamma ^2 
\hat{\tau}_1^2 v^2+1)\hat{D}(\tau_1)}{\gamma  v^2}\Big]-\frac{(2 \gamma 
^2-1) a_2\cdot \hat{b}\, a_2\cdot v_1 (\gamma ^2 \hat{\tau}_1^2 
v^2+1)\hat{D}(\tau_1)}{\gamma ^2 v^2}\bigg]\, .
   \end{aligned}&&
   \end{flalign}
   
   \begin{flalign}
   \begin{aligned}
   &F_2= \frac{m_2}{\hat{D}(\tau_1)^3 |b|^2}\bigg[a_1\cdot \hat{b} \Big[-\gamma \hat{\tau}_1 (2 
(\gamma v)^2 \hat{\tau}_1^2+3) a_2\cdot \hat{b}-\frac{a_2\cdot 
v_1}{\gamma  v^2}\Big]+a_1\cdot v_2 \Big[\frac{a_2\cdot \hat{b}}{\gamma ^2 
v^2}+\hat{\tau}_1 (\gamma ^2 \hat{\tau}_1^2+1) a_2\cdot v_1\Big]\\
&- a_1\cdot a_2\,\gamma  
\hat{\tau}_1 [(\gamma v)^2 \hat{\tau}_1^2+1]-a_2^2 \gamma  
\hat{\tau}_1[(\gamma v)^2 \hat{\tau}_1^2+1]-\gamma  \hat{\tau}_1 [2 
(\gamma v)^2 \hat{\tau}_1^2+3] (a_2\cdot \hat{b})^2-\frac{2 a_2\cdot 
\hat{b}\, a_2\cdot v_1}{\gamma  v^2}-\gamma \hat{\tau}_1^3 
(a_2\cdot v_1)^2\\
&+ \frac{(\gamma\,a_1\cdot \hat{b}\, a_2\cdot v_1-a_1\cdot v_2\, a_2\cdot \hat{b}) [\gamma ^2 \hat{\tau}_1^2 
v^2+1]\hat{D}(\tau_1)}{\gamma^2  v^2}+\frac{a_2\cdot \hat{b}\, a_2\cdot 
v_1 [2 \gamma ^2 \hat{\tau}_1^2 v^2+2]\hat{D}(\tau_1)}{\gamma  v^2}\bigg]\, .
   %G m_2\bigg[-\frac{\gamma   \bar{\bar{\tau}} \big[a_1\cdot 
   %a_2+a_2^2\big]}{(\gamma ^2-1) \hat{D}(\bar{\bar{\tau}}) |b|^2}-\frac{ a_1\cdot v_2 \big[(\hat{D}(\bar{\bar{\tau}})^3-1) a_2
  % \cdot \hat{b}-\bar{\bar{\tau}} (\frac{\gamma ^2 
   %\bar{\bar{\tau}}^2}{(\gamma ^2-1)^2}+1) a_2\cdot v_1\big]}{(\gamma 
   %^2-1) \hat{D}(\bar{\bar{\tau}})^3 |b|^2}\\
  % &+\frac{\gamma a_1\cdot \hat{b} 
   %\big[(\hat{D}(\bar{\bar{\tau}})^3-1) a_2\cdot v_1-(2 \hat{D}(\bar{
   %\bar{\tau}})^2 \bar{\bar{\tau}}+\bar{\bar{\tau}}) a_2\cdot 
   %\hat{b}\big]}{(\gamma ^2-1) \hat{D}(\bar{\bar{\tau}})^3 |b|^2}\\
   %&+\frac{\gamma a_2\cdot v_1 \big[2 (\gamma ^2-1)^2 
   %(\hat{D}(\bar{\bar{\tau}})^3-1) a_2\cdot 
   %\hat{b}-\bar{\bar{\tau}}^3 a_2\cdot v_1\big]}{(\gamma ^2-1)^3 
   %\hat{D}(\bar{\bar{\tau}})^3 |b|^2}\\
   %&-\frac{(a_2\cdot \hat{b})^2\gamma  \big[2 \hat{D}(\bar{\bar{\tau}})^2+1\big] 
   %\bar{\bar{\tau}}}{(\gamma ^2-1) 
   %\hat{D}(\bar{\bar{\tau}})^3 |b|^2}\bigg]\ .
   \end{aligned}&&
   \end{flalign}
with $\hat{D}(\tau)$ introduced in the main text~\eqref{Dtau}.
   For a better overview, we split $F_3$ further: $F_3=\sum_{i=1}^5 F_{3 i}$.

   \begin{flalign}
   \begin{aligned}
   &F_{31}= \frac{m_2}{(\gamma v)^4|b|^2} \bigg[a_1^2 \Big[\frac{3 \gamma ^2-2 (2 \gamma ^4-3 
\gamma ^2+1) \gamma ^4 \hat{\tau}_1^4 v^4-3 (\gamma v)^4 (2 
\gamma ^2-1) \hat{\tau}_1^2+\gamma ^4 (v^4-2)-1}{2  
\hat{D}(\tau_1)^3}\\
&-\frac{3 \gamma ^2+2 (2 \gamma 
^4-3 \gamma ^2+1) \gamma  \hat{\tau}_1 v+\gamma ^4 (v^4-2)-1}{2 
}\Big]\\
&+a_1\cdot a_2 \Big[-\frac{(2 \gamma ^4-3 \gamma 
^2+1) (2 \gamma ^4 \hat{\tau}_1^4 v^4+3 (\gamma v)^2 
\hat{\tau}_1^2+1)+(2 \gamma ^4-3 \gamma ^2+1) (2 \gamma  \hat{\tau}_1 
v-1) \hat{D}(\tau_1)^3}{ \hat{D}(\tau_1)^3
}\Big]\\
&+a_2^2 \Big[\frac{-3 \gamma ^2-2 (2 \gamma 
^4-3 \gamma ^2+1) \gamma  \hat{\tau}_1 v+2 \gamma ^4 (v^4+1)+1}{2 
}\\
&-\frac{-3 \gamma ^2+2 (2 \gamma ^4-3 \gamma 
^2+1) \gamma ^4 \hat{\tau}_1^4 v^4+3 (\gamma v)^4 (2 \gamma 
^2-1) \hat{\tau}_1^2+2 \gamma ^4 (v^4+1)+1}{2 
\hat{D}(\tau_1)^3}\Big]\bigg]\, ,
   %\frac{G m_2}{2 |b|^2}\bigg[a_1^2 \bigg[\frac{\gamma ^2}{\gamma 
   %^2-1}+\frac{-\gamma ^4+\gamma ^2+(1-2 \gamma ^2) 
   %\bar{\bar{\tau}}^2}{(\gamma ^2-1)^2 
   % \hat{D}(\bar{\bar{\tau}})^3}+\frac{2 (1-2 \gamma ^2) 
   % \bar{\bar{\tau}}^2}{(\gamma ^2-1)^2 
   % \hat{D}(\bar{\bar{\tau}})}+\frac{(2-4 \gamma ^2) 
   % \bar{\bar{\tau}}}{(\gamma ^2-1)^{3/2}}\bigg]\\
   % &-\frac{2\, a_1\cdot a_2\, (2 \gamma ^2-1) 
   % \big[-\sqrt{\gamma ^2-1}+\sqrt{\gamma ^2-1} 
   % \hat{D}(\bar{\bar{\tau}}) (2 \hat{D}(\bar{\bar{\tau}})-1)+2 
   % \hat{D}(\bar{\bar{\tau}}) \bar{\bar{\tau}}\big]}{(\gamma 
   % ^2-1)^{3/2} \hat{D}(\bar{\bar{\tau}})}\\
   % &-a_2^2 \bigg[\frac{2 
   % \bar{\bar{\tau}}^4}{(\gamma ^2-1)^3 
   % \hat{D}(\bar{\bar{\tau}})^3}-\frac{-2 \gamma ^2 
   % ((\hat{D}(\bar{\bar{\tau}})-3) 
   % \hat{D}(\bar{\bar{\tau}})^2+2)+3 
   % (\hat{D}(\bar{\bar{\tau}})-2) 
   % \hat{D}(\bar{\bar{\tau}})^2+3}{(\gamma ^2-1) 
   % \hat{D}(\bar{\bar{\tau}})^3}\\
   % &-\frac{3 \bar{\bar{\tau}}^2 (4 
   % (\gamma ^2-1) \hat{D}(\bar{\bar{\tau}})^2-1)}{(\gamma ^2-1)^2 
   % \hat{D}(\bar{\bar{\tau}})^3}-\frac{2 (6 \gamma ^2-7) 
   % \bar{\bar{\tau}}}{(\gamma ^2-1)^{3/2}}\bigg]\bigg]\ ,
   \end{aligned}&&
\end{flalign}
\begin{flalign}
\begin{aligned}
   &F_{32}= \frac{m_2\, a_1\cdot v_2 }{v^2|b|^2}\bigg[\frac{
\hat{\tau}_1 ((\gamma v)^2 \hat{\tau}_1^2+2) a_1\cdot 
\hat{b}}{\gamma \hat{D}(\tau_1)^3}+\frac{\hat{\tau}_1 ((\gamma v)^2 
\hat{\tau}_1^2+2) a_2\cdot \hat{b}}{\gamma
\hat{D}(\tau_1)^3}\\
&+a_2\cdot v_1 \Big[\frac{2 (\gamma v)^4 
(2 \gamma ^2-1) \hat{\tau}_1^4+(6 \gamma ^4-9 \gamma ^2+3) 
\hat{\tau}_1^2+2\gamma^2}{\gamma^3 v^2\hat{D}(\tau_1)^3}+\frac{2 
((\gamma v)^2 \hat{\tau}_1^2+1)}{\gamma 
\hat{D}(\tau_1)^2}+\frac{2 (2 \gamma ^2-1) 
(\gamma  \hat{\tau}_1 v-1)}{\gamma ^3  v^2}\Big]\\
%squared
&+ a_1\cdot v_2 \Big[\frac{-5 \gamma ^2+2 (1-2 \gamma 
^2) \gamma ^4 \hat{\tau}_1^4 v^4+(-6 \gamma ^4+9 \gamma ^2-3) \hat{\tau}_1^2-2
 \gamma ^4 (v^4-1)+1}{2 \gamma ^4 \hat{D}(\tau_1)^3
v^2}-\frac{2  ((\gamma v)^2 \hat{\tau}_1^2+1)}{ 
\hat{D}(\tau_1)^2 }\\
&+\frac{\gamma ^2+(2 \gamma -4 \gamma 
^3) \hat{\tau}_1 v+2 \gamma ^4 (v^4+1)-1}{2 \gamma ^4 v^2}\Big]\bigg]\, ,
   %\frac{G m_2\, a_1\cdot v_2}{2 (\gamma 
   % ^2-1)^{7/2} \hat{D}(\bar{\bar{\tau}})^3 |b|^2}\bigg[2\,a_1\cdot 
   % \hat{b}\, \gamma  \sqrt{\gamma ^2-1} 
   % \bar{\bar{\tau}} \big[2 \gamma ^2+\bar{\bar{\tau}}^2-2\big]\\
   % &+a_1\cdot v_2 \big[(\gamma ^2-1) 
   % \hat{D}(\bar{\bar{\tau}})^3 (\sqrt{\gamma ^2-1} (\gamma 
   % ^2+1)+(2-4 \gamma ^2) \bar{\bar{\tau}})+2 (1-2 \gamma ^2) 
   % \sqrt{\gamma ^2-1} \hat{D}(\bar{\bar{\tau}})^2 
   % \bar{\bar{\tau}}^2+\sqrt{\gamma ^2-1} (-\gamma ^4+(1-2 \gamma 
   % ^2) \bar{\bar{\tau}}^2+1)\big]\\
   % &+2\, a_2\cdot 
   % \hat{b}\, \gamma  (\gamma ^2-1)^{3/2} 
   % \big[\hat{D}(\bar{\bar{\tau}})^2+1\big] \bar{\bar{\tau}}\\
   % &+2\,a_2\cdot v_1\,\gamma\big[(-2 (\gamma ^2-1) 
   % \hat{D}(\bar{\bar{\tau}})^3 (\sqrt{\gamma ^2-1} \gamma ^2-2 
   % \gamma ^2 \bar{\bar{\tau}}+\bar{\bar{\tau}})+2 \sqrt{\gamma 
   % ^2-1} (2 \gamma ^2-1) \hat{D}(\bar{\bar{\tau}})^2 \
   % \bar{\bar{\tau}}^2\\
   % &+\sqrt{\gamma ^2-1} (2 \gamma ^2 (\gamma 
   % ^2-1)+(2 \gamma ^2-1) \bar{\bar{\tau}}^2))\big]\bigg]\ ,
\end{aligned}&&
\end{flalign}
\begin{flalign}
\begin{aligned}
   &F_{33}=\frac{m_2\, a_1\cdot \hat{b}}{(\gamma v)^4|b|^2} \bigg[a_2\cdot \hat{b} 
\Big[-3 \gamma ^2-4 (2 \gamma ^4-3 \gamma ^2+1) \gamma  \hat{
\tau} v+\gamma ^4 (v^4+2)+1\\
&-\frac{-3 
\gamma ^2+4 (2 \gamma ^4-3 \gamma ^2+1) \gamma ^4 \hat{\tau}_1^4 
v^4+6 (\gamma v)^4 (2 \gamma ^2-1) \hat{\tau}_1^2+\gamma ^4 
(v^4+2)+1}{ \hat{D}(\tau_1)^3}\Big]\\
&+\frac{\hat{\tau}_1 a_2\cdot v_1 [6 \gamma ^2-(\gamma 
^2-1)^2 (2 \gamma ^2-1) \hat{\tau}_1^2+\gamma ^4 \
(v^4-4)-2]}{ \hat{D}(\tau_1)^3 
}\\
&-a_1\cdot \hat{b} \Big[\frac{(2 \gamma ^4-3 \gamma ^2+1) 
(4 \gamma ^4 \hat{\tau}_1^4 v^4+6 (\gamma v)^2 
\hat{\tau}_1^2+1)}{2  \hat{D}(\tau_1)^3}
+\frac{(2 \gamma ^4-3 \gamma ^2+1) (4 \gamma  \hat{\tau}_1 
v-1)}{2}\Big]\bigg]\, ,
   % \frac{G m_2\,  a_1\cdot \hat{b}}{2 
   % \hat{D}(\bar{\bar{\tau}})^3 |b|^2}\bigg[\frac{2}{\gamma 
   % ^2-1} \bigg(a_2\cdot b \bigg[-3 \gamma 
   % ^2+\hat{D}(\bar{\bar{\tau}})^3 \bigg(\gamma ^2 \bigg(3-\frac{8 
   % \bar{\bar{\tau}}}{\sqrt{\gamma ^2-1}}\bigg)+\frac{4 
   % \bar{\bar{\tau}}}{\sqrt{\gamma ^2-1}}-2\bigg)-\frac{2 (2 \gamma 
   % ^2-1) (2 \hat{D}(\bar{\bar{\tau}})^2+1) 
   % \bar{\bar{\tau}}^2}{\gamma ^2-1}+2\bigg]\\
   % &+a_2\cdot 
   % v_1\,\bar{\bar{\tau}} \bigg[-\hat{D}(\bar{\bar{\tau}})^2-\frac{\gamma ^2 (2 \gamma 
   % ^2+\bar{\bar{\tau}}^2-2)}{(\gamma ^2-1)^2}\bigg]\bigg)\\
   % &-\frac{a_1\cdot \hat{b}\,(2 \gamma ^2-1) \big[\sqrt{\gamma 
   % ^2-1} (2 (\hat{D}(\bar{\bar{\tau}})^2-1) (\frac{2 
   % \bar{\bar{\tau}}^2}{\gamma 
   % ^2-1}+3)+1)-\hat{D}(\bar{\bar{\tau}})^3 (\sqrt{\gamma ^2-1}-4 
   % \bar{\bar{\tau}})\big]}{(\gamma ^2-1)^{3/2}}\bigg]\ ,
\end{aligned}&&
\end{flalign}
\begin{flalign}
\begin{aligned}
   &F_{34}=\frac{m_2\, a_2\cdot 
v_1}{(\gamma v)^4|b|^2}\bigg[\frac{\hat{\tau}_1 a_2\cdot \hat{b} [6
\gamma ^2-(\gamma v)^4 (2 \gamma ^2-1) \hat{\tau}_1^2+2 \gamma ^4 
(v^4-2)-2]}{ \hat{D}(\tau_1)^3}\\
&-a_2\cdot 
v_1 \Big[\frac{(2 \gamma ^2-1) (2 \gamma ^4 \hat{\tau}_1^4 v^4+3 
(\gamma v)^2 \hat{\tau}_1^2+2)+2(2 \gamma ^2-1) (\gamma  \hat{\tau}_1 v-1)\hat{D}(\tau_1)^3}{2 \hat{D}(\tau_1)^3 
}\Big]\bigg]\, ,
   %-\frac{G m_2\, a_2\cdot v_1}{2 |b|^2} \bigg[ a_2\cdot v_1 \bigg[\frac{2 (\gamma 
   % ^2-1)^2 (2 \gamma ^2-1)+2 \bar{\bar{\tau}}^4+3 (\gamma ^2-1) 
   % \bar{\bar{\tau}}^2}{(\gamma ^2-1)^4 
   % \hat{D}(\bar{\bar{\tau}})^3}-\frac{6 (\gamma ^2+2 
   % \bar{\bar{\tau}}^2-1)}{(\gamma ^2-1)^2 
   % \hat{D}(\bar{\bar{\tau}})}\\
   % &+\frac{2 ((\gamma ^2-2) 
   % \sqrt{\gamma ^2-1}+(7-6 \gamma ^2) \bar{\bar{\tau}})}{(\gamma 
   % ^2-1)^{5/2}}\bigg]+\frac{2\,  a_2\cdot \hat{b}\, \bar{\bar{\tau}}\big[(2 \gamma ^2-1) 
   % \hat{D}(\bar{\bar{\tau}})^2+1\big]}{(\gamma 
   % ^2-1)^2 \hat{D}(\bar{\bar{\tau}})^3}\bigg]\ ,
\end{aligned}&&
\end{flalign}
\begin{flalign}
\begin{aligned}
   &F_{35}=\frac{ m_2\, (a_2\cdot \hat{b})^2}{(\gamma v)^4|b|^2} \bigg[\frac{-3 \gamma ^2-4
(2 \gamma ^4-3 \gamma ^2+1) \gamma  \hat{\tau}_1 v+\gamma ^4 (4 
v^4+2)+1}{2}\\
&-\frac{-3 \gamma ^2+4 (2 \gamma 
^4-3 \gamma ^2+1) \gamma ^4 \hat{\tau}_1^4 v^4+6 (\gamma v)^4 
(2 \gamma ^2-1) \hat{\tau}_1^2+\gamma ^4 (4 v^4+2)+1}{2 
\hat{D}(\tau_1)^3}\bigg]\, .
   % G m_2 (a_2\cdot b)^2\bigg[\frac{ (8 (\gamma ^2-1)^{7/2} \hat{D}(\bar{
   % \bar{\tau}})^4+\hat{D}(\bar{\bar{\tau}})^3 ((\gamma 
   % ^2-1)^{5/2}+4 (2 \gamma ^2-3) (\gamma ^2-1)^2 
   % \bar{\bar{\tau}}))}{2 (\gamma 
   % ^2-1)^{7/2} \hat{D}(\bar{\bar{\tau}})^3 |b|^2}\\
   % &+\frac{-2 \sqrt{\gamma ^2-1} (2 \gamma ^6-3 \gamma 
   % ^4+1) \hat{D}(\bar{\bar{\tau}})^2+\sqrt{\gamma ^2-1} (-(\gamma 
   % ^2-1)^2 (4 \gamma ^2-9)-4 \bar{\bar{\tau}}^4)}{2 (\gamma 
   % ^2-1)^{7/2} \hat{D}(\bar{\bar{\tau}})^3 |b|^2}\bigg]\ ,
   \end{aligned}&&
   \end{flalign}
   \begin{equation}
   \begin{split}
   &F_4=\frac{ m_2}{\hat{D}(\hat{\tau_1})^3 |b|^2} \bigg[\frac{(2 \gamma ^2-1) a_1\cdot \hat{b}\, (
\gamma ^2 \hat{\tau_1}^2 v^2+1) (2 \gamma ^2 \hat{\tau_1}^2 v^2+1) 
(a_1\cdot \hat{\ell}+a_2\cdot \hat{\ell})}{\gamma ^2 v^2}\\
&+\frac{a_2\cdot \hat{b}\, 
[a_1\cdot \hat{\ell} (\gamma ^2+2 (2 \gamma ^2-1) \gamma ^4 \hat{\tau_1}^4 
v^4+(6 \gamma ^4-9 \gamma ^2+3) \hat{\tau_1}^2)+a_2\cdot \hat{\ell} (2 (2 
\gamma ^2-1) \gamma ^4 \hat{\tau_1}^4 v^4+(6 \gamma ^4-9 \gamma 
^2+3) \hat{\tau_1}^2+1)]}{\gamma ^2 v^2}\\
&+\frac{\hat{\tau_1} 
a_2\cdot v_1 [(3 \gamma ^2+(2 \gamma ^4-3 \gamma ^2+1) 
\hat{\tau_1}^2-2) a_1\cdot \hat{\ell}+(4 \gamma ^2+(2 \gamma ^4-3 \gamma 
^2+1) \hat{\tau_1}^2-3) a_2\cdot \hat{\ell}]}{\gamma ^2 
v^2}\\
&-\frac{\hat{\tau_1} a_1\cdot v_2 (\gamma ^2 \hat{\tau_1}^2 
v^2+1) (a_1\cdot \hat{\ell}+a_2\cdot \hat{\ell})}{\gamma  
v^2}+ \frac{(1-2 
\gamma ^2) a_1\cdot \hat{b} (1-2 \gamma  \hat{\tau_1} v) (\gamma 
^2 \hat{\tau_1}^2 v^2+1) (a_1\cdot \hat{\ell}+a_2\cdot \hat{\ell})\hat{D}(\tau_1)}{\gamma ^2 
v^2}\\
&+\frac{a_2\cdot \hat{b}\, (\gamma ^2 \hat{\tau_1}^2 v^2+1) 
[\gamma  a_1\cdot \hat{\ell} (2 (2 \gamma ^2-1) \hat{\tau_1} v-\gamma 
)+a_2\cdot \hat{\ell} (4 \gamma ^3 \hat{\tau_1} v-2 \gamma  \hat{\tau_1}
v-1)]\hat{D}(\tau_1)}{\gamma ^2 v^2}\bigg]\, .
   % \frac{G m_2}{|b|^2} \Bigg[\frac{a_1\cdot \hat{b}\,(2 \gamma ^2-1) \big[2 
   % \hat{D}(\bar{\bar{\tau}})^2+\hat{D}(\bar{\bar{\tau}}) (\frac{2 
   % \bar{\bar{\tau}}}{\sqrt{\gamma ^2-1}}-1)-1\big] (a_1\cdot L+S_2\cdot 
   % L)}{(\gamma ^2-1) \hat{D}(\bar{\bar{\tau}})}\\
   % &+S_2\cdot \hat{b} \bigg[\frac{
   % \bar{\bar{\tau}}^2 (\gamma ^2+(4 \gamma ^2-2) \hat{D}(\bar{\bar{\tau}})-1) (a_1\cdot L+S_2\cdot L)}{(\gamma ^2-1)^2 \hat{D}(\bar{\bar{\tau}})^2}
   % -\frac{(\hat{D}(\bar{\bar{\tau}})-1) (-\gamma ^2+(2 \gamma 
   % ^2-1) \hat{D}(\bar{\bar{\tau}})^2+1) (a_1\cdot L+S_2\cdot L)}{(\gamma ^2-1) 
   % \hat{D}(\bar{\bar{\tau}})^3}\\
   % &+\frac{2 (2 \gamma ^2-1) \bar{\bar{\tau}} 
   % (a_1\cdot L+S_2\cdot L)}{(\gamma ^2-1)^{3/2}}+\bigg(\frac{\hat{D}(\bar{\bar{\tau}})-1}{\hat{D}
   % (\bar{\bar{\tau}})^3}+\frac{\bar{\bar{\tau}}^2}{(\gamma ^2-1) \hat{D}(\bar{\bar{\tau}})^2}\bigg)S_2\cdot L\bigg]\\
   % &+\frac{S_2\cdot v_1\,\bar{\bar{\tau}} \big[a_1\cdot L (\gamma 
   % ^2 (\hat{D}(\bar{\bar{\tau}})^2+2)+\bar{\bar{\tau}}^2-2)+(2 (\gamma 
   % ^2-1) (\hat{D}(\bar{\bar{\tau}})^2+1)+\hat{D}(\bar{\bar{\tau}})^2) 
   % S_2\cdot L\big]}{(\gamma ^2-1)^2 \hat{D}(\bar{\bar{\tau}})^3}
   % \\
   % &-\frac{a_1\cdot v_2\,\gamma  \bar{\bar{\tau}} \big[a_1\cdot 
   % L+S_2\cdot L\big]}{(\gamma ^2-1)^2 \hat{D}(\bar{\bar{\tau}})}\Bigg]\ .
   \end{split}
   \end{equation}
In the results displayed above, we have introduced the dimensionless variable $\hat{\tau}_1:=\tau_1/|b|$.

\section{1PM Fourier integrals}\label{sec:1PMfourier}

Here we discuss the Fourier integrals~\eqref{1PMfamily}
which were used in the 1PM results in Section~\ref{LeadingTrajectory}.
The family is:
\begin{equation}
\label{1pmIntegrals}
I^{\mu_1...\mu_n}_{\alpha,\,\beta}=\int_q\frac{\deltabar(q\cdot v_2)}{(q^2)^\alpha (q\cdot v_1+\i 0)^\beta}e^{\i q\cdot\tilde{b}}q^{\mu_1}\cdots q^{\mu_n}\, .
\end{equation}
We start by analyzing the sub-family $I_{\alpha,0}$ that does not include worldline propagators.
% Every integral in this family can be derived from the simpler family $I_{\alpha,0}$:
% \begin{equation}
% I_{\alpha,0}=\int_{q}\frac{\deltabar(q\cdot v_2)}{(q^2)^{\alpha}}\,e^{iq\cdot \tilde{b}}\ .
% \end{equation}
The delta function is resolved by going into the rest frame of BH2,
and the resulting integral is solved by Schwinger parameterization.
The covariant solution is given below,
where $P_2^\mn$ projects onto the rest frame of BH2:
\begin{align}
I_{\alpha,0}&=\frac{(-1)^\alpha\Gamma(\frac{D-1}{2}-\alpha)}{4^\alpha\sqrt{\pi}^{D-1} \Gamma(\alpha) }
\left(-\tilde{b}\cdot P_2\cdot \tilde{b}\right)^{\alpha-\frac{D-1}{2}}\,, &
P_2^\mn&:=\eta^\mn-v_2^\mu v_2^\nu\,.
\end{align}
We generalize to $\beta\neq0$, i.e.~introduce retarded worldline propagators, using
\begin{equation}
\label{WLpropagatorTrick}
\int_\omega e^{-\i\omega\tau}\frac{f(\omega)}{(\omega+\i 0)^\beta}=
(-1)^{\beta/2}\int_{-\infty}^\tau \!\d\tau_1\cdots\int_{-\infty}^{\tau_{\beta-2}} \!\d\tau_{\beta-1} \int_{-\infty}^{\tau_{\beta-1}}\!\d\tau_\beta\int_\omega e^{\i\omega\tau_\beta}f(\omega)\,.
\end{equation}
Powers of the worldline propagator are added by successive integration over the time parameter $\tau$.
Integrals of higher tensor rank $I^{\mu_1...\mu_n}_{\alpha,\,\beta}$
are then obtained by differentiating with respect to the \emph{unconstrained} $b^\mu$:
\begin{equation}
I^{\mu_1...\mu_n}_{\alpha,\,\beta}=
(-\i)^{n}\frac{\partial}{\partial{b}_{(\mu_n}}\cdots\frac{\partial}{\partial{b}_{\mu_1)}}I_{\alpha,\,\beta}\ .
\end{equation}
One does not impose $b\cdot v_i=0$ before taking derivatives. 

\section{One-loop integrals}\label{loopIntegration}

In order to compute 2PM trajectories we introduced the following pair of one-loop integral families,
in both cases with $q\cdot v_2=0$:
\begin{subequations}
\begin{align}
I^{1-\mathrm{loop}}_{\nu_1,\nu_2,\nu_3}&=
\int_\ell\frac{\deltabar(\ell\cdot v_2)}{(\ell^2)^{\nu_1}((\ell-q)^2)^{\nu_2}(\ell\cdot v_1+\i 0)^{\nu_3}}\ ,\\
J^{1-\mathrm{loop}}_{\nu_1,\nu_2,\nu_3}&=
\int_\ell\frac{\deltabar(\ell\cdot v_1)}{(\ell^2)^{\nu_1}((\ell-q)^2+\i 0\ \mathrm{sgn}(\ell^0-q^0))^{\nu_2}(\ell\cdot v_2+\i 0)^{\nu_3}}\ .
\end{align}
\end{subequations}
Both integral families only depend on $\{|q|, \gamma, q\cdot v_1\}$.
Introducing $\hat{q}\cdot v_1:=q\cdot v_1/|q|$ instead of $q\cdot v_1$
leaves $|q|$ as the only dimensionful quantity,
and so its dependence in the integrals is simply determined by counting overall mass dimensions.
This leaves two dimensionless parameters $X=\{\gamma,\hat{q}\cdot v_1\}$.
As stated in the main text, the two integral families are spanned by the following masters:
\begin{align}\label{masterInts}
   \vec{I}&=(I^{1-\mathrm{loop}}_{0,1,1}, I^{1-\mathrm{loop}}_{1,1,0}, I^{1-\mathrm{loop}}_{1,1,1})\,, &
   \qquad \vec{J}&=(J^{1-\mathrm{loop}}_{0,1,0}, J^{1-\mathrm{loop}}_{0,1,1}, J^{1-\mathrm{loop}}_{1,1,0}, J^{1-\mathrm{loop}}_{1,1,1})\,.
\end{align}
Final expressions for the $I$-integrals~\eqref{eq:1loopI} and $J$-integrals~\eqref{eq:1loopJ}
were given in the main text.
In this Appendix we outline our procedure for deriving these expressions,
using integration-by-parts (IBPs) reduction~\cite{Maierhofer:2017gsa,Klappert:2020nbg} to master integrals,
differential equations involving $\hat{q}\cdot v_1$
and the method of regions for boundary integrals in the limit $\hat{q}\cdot v_1\to0$.
This corresponds with the asymptotic limit $\tau_i\to\pm\infty$.

\subsubsection{Differential equations}

Although the present integral families
involve two parameters, it suffices to consider differential
equations only in $\hat{q}\cdot v_1$ as all $\gamma$-dependence is fixed by the boundary integrals.
In the first family $I^{1-\mathrm{loop}}_{\nu_1,\nu_2,\nu_3}$
we find it a natural choice to consider the differential equation with respect to
$(q\cdot v_1)_N:=\hat{q}\cdot v_1/(\gamma v)$,
as this allows us to pull all $\gamma$-dependence out of the integrals.
This dependence is then given by an overall factor of $1/(\gamma v)^{\nu_3}$ ---
in short, the dependence on $\gamma$ is made trivial by working in the frame $v_2^\mu=(1,\mathbf{0})$.
The two differential equations are then given by (where we set $|q|=1$ since its dependence is trivially determined by the mass scale of the integral):
\begin{subequations}
\begin{align}
   \frac{\mathrm{d}\,\vec{I}}{\mathrm{d}\,(q\cdot v_1)_N}&=A_I \vec{I}\, , &
   A_I=2 \eps
   \begin{pmatrix}
   -\frac{1}{(q\cdot v_1)_N} & 0 & 0\\
   0 & 0 &0\\
   -\frac{1}{(q\cdot v_1)_N((q\cdot v_1)_N^2-1)} & -\frac{2}{((q\cdot v_1)_N^2-1) \gamma v} &
   \frac{(q\cdot v_1)_N}{(q\cdot v_1)_N^2-1}
   \end{pmatrix}\,,\\
   \frac{\mathrm{d}\,\vec{J}}{\mathrm{d}\,\hat{q}\cdot v_1}&=A_J \vec{J}\, , &
   A_J=
   \begin{pmatrix}
   \frac{1-2\eps}{\hat{q}\cdot v_1} & 0 & 0 & 0\\
   0 & -\frac{2\eps}{\hat{q}\cdot v_1} & 0 & 0\\
   \frac{2\eps-1}{\hat{q}\cdot v_1(1+\hat{q}\cdot v_1^2)} & 0 & \frac{\hat{q}\cdot v_1(2\eps-1)}{1+\hat{q}\cdot v_1^2} & 0\\
   0 & -\frac{2\eps(\gamma^2-1)}{\hat{q}\cdot v_1(1+\hat{q}\cdot v_1^2-\gamma^2)} &
   \frac{4 \gamma \eps}{1+\hat{q}\cdot v_1^2-\gamma^2} & \frac{2\hat{q}\cdot v_1 \eps}{1+\hat{q}\cdot v_1^2-\gamma^2}
   \end{pmatrix}\,.
\end{align}
\end{subequations}
% \begin{equation}
% \frac{\mathrm{d}\,\vec{I}}{\mathrm{d}\,(q\cdot v_1)_N}=A_I \vec{I}\, ,\qquad \frac{\mathrm{d}\,\vec{J}}{\mathrm{d}\,\hat{q}\cdot v_1}=A_J \vec{J}\, ,
% \end{equation}
% \begin{equation}
% A_I=\eps
% \begin{pmatrix}
% -\frac{1}{(q\cdot v_1)_N} & 0 & 0\\
% 0 & 0 &0\\
% -\frac{1}{(q\cdot v_1)_N((q\cdot v_1)_N^2-1)} & -\frac{2}{((q\cdot v_1)_N^2-1) \gamma v} &
%  \frac{(q\cdot v_1)_N}{(q\cdot v_1)_N^2-1}
% \end{pmatrix}\ ,
% \end{equation}
% \begin{equation}
% A_J=
% \begin{pmatrix}
% \frac{1-\eps}{\hat{q}\cdot v_1} & 0 & 0 & 0\\
% 0 & -\frac{\eps}{\hat{q}\cdot v_1} & 0 & 0\\
% \frac{\eps-1}{\hat{q}\cdot v_1(1+\hat{q}\cdot v_1^2)} & 0 & \frac{\hat{q}\cdot v_1(\eps-1)}{1+\hat{q}\cdot v_1^2} & 0\\
% 0 & -\frac{\eps(\gamma^2-1)}{\hat{q}\cdot v_1(1+\hat{q}\cdot v_1^2-\gamma^2)} &
%  \frac{2 \gamma \eps}{1+\hat{q}\cdot v_1^2-\gamma^2} & \frac{\hat{q}\cdot v_1 \eps}{1+\hat{q}\cdot v_1^2-\gamma^2}
% \end{pmatrix}\ .
% \end{equation}
While the first DE is already in $\eps$-form,
we require a transformation in order to bring the
differential equation represented by $A_J$ into $\eps$-form.
We therefore act with
\begin{equation}
T_J=
\begin{pmatrix}
-\frac{2\eps-1}{2\eps \hat{q}\cdot v_1} & 0 & 0 & 0\\
0 & 1 & 0 & 0\\
0 & 0 & \sqrt{(\hat{q}\cdot v_1)^2+1} & 0\\
0 & 0 & 0 & 1
\end{pmatrix}
\end{equation}
on the basis vector $\vec{J}$.
Expressions for the master integrals are then built up order-by-order in $\eps$,
and our next task is to fix the corresponding constants of integration.

\subsubsection{Evaluation of boundary integrals}

In order to fix boundary conditions on our master integrals we consider the limit $q\cdot v_1\to0$,
wherein the integrals depend non-trivially only on $\gamma$.
We claim that there are two regions involved, 
characterised by the following scalings of the loop momentum $\ell$:
\begin{align}\label{mspaceRegions}
\ell^\mu &\sim (q\cdot v_1, 1)\,, &  \ell^\mu &\sim (q\cdot v_1 -\i 0 ,q\cdot v_1 - \i 0)\,.
\end{align}
We refer to the latter as the \emph{soft} region.
%\cite{DiVecchia:2022owy}
We will justify this claim in the next subsection;
here, we focus on evaluating the corresponding boundary integrals.
For the purposes of this discussion, we focus on the second (and more non-trivial)
integral family $J^{1-\mathrm{loop}}_{\nu_1,\nu_2,\nu_3}$.
Boundary integrals for the family $I^{1-\mathrm{loop}}_{\nu_1,\nu_2,\nu_3}$
are obtained by a similar procedure.

To illustrate our method, consider the specific master integral $J^{1-\mathrm{loop}}_{1,1,1}$.
Working in the inertial frame $v_1^\mu=(1,\mathbf{0})$, $v_2^\mu=\gamma(1,\mathbf{v})$
the integral is massaged into a (for our purposes) more convenient form:
\begin{equation}
\label{miJ111}
\begin{split}
J^{1-\mathrm{loop}}_{1,1,1}&=\int_\ell\frac{\deltabar(\ell\cdot v_1)}{\ell^2\,((\ell-q)^2+\text{sgn}(\ell^0-q^0)\i 0)\, (\ell\cdot v_2+\i 0)}\\
&=\int_{\mathbf{l}} \frac{1}{(-\mathbf{l}^2)\,
((q\cdot v_1)^2-(\mathbf{l}-\mathbf{q})^2-\text{sgn}(q\cdot v_1)\i 0)\,
(-\gamma(\mathbf{l}\cdot \mathbf{v}_2)+\i 0)}\\
&=\frac{1}{\gamma}\int_{\mathbf{l}} \frac{1}{(\mathbf{l}+\mathbf{q})^2\, (-(q\cdot v_1-\i0)^2+\mathbf{l}^2)\, (-(q\cdot v_1-\i0)-\mathbf{l}\cdot \mathbf{v})}\ .
\end{split}
\end{equation}
having used $q\cdot v_2=0$, which implies that $\mathbf{q}\cdot\mathbf{v}=q\cdot v_1$.
Assuming the two scalings~\eqref{mspaceRegions}
and expanding to leading order in $q\cdot v_1$,
one arrives at (discarding the factor of  $\gamma$ in the denominator)
\begin{subequations}
\begin{align}
&J^{1-\mathrm{loop}}_{1,1,1}\xrightarrow{q\cdot v_1\to0}-\int_\mathbf{l}\frac{1}{(\mathbf{l}+\mathbf{q})^2\mathbf{l}^2 (\mathbf{l}\cdot \mathbf{v})}\,, &
&\ell^\mu \sim (q\cdot v_1, 1)\,\label{J111reg1}\\
&J^{1-\mathrm{loop}}_{1,1,1}\xrightarrow{q\cdot v_1\to0}-(q\cdot v_1\!-\!\i0)^{D-4}\frac{1}{\mathbf{q}^2}\int_\mathbf{l}\frac{1}{(\mathbf{l}^2-1)(1+\mathbf{l}\cdot \mathbf{v} -\i 0)}\,, & 
&\ell^\mu \sim (q\cdot v_1, q\cdot v_1)\,.\label{J111regzf}
\end{align}
\end{subequations}
To arrive at the integrand in the latter soft region
we rescaled $\mathbf{l}\to q\cdot v_1 \mathbf{l}$.
Hence, $\mathbf{l}$ is to be considered dimensionless in the lowermost of the two integrands.
Computing these integrals yields the boundary conditions.

The first region $\ell^\mu \sim (q\cdot v_1, 1)$ is more straightforward to compute,
as it corresponds simply with the one-loop integral family used
in calculations of asymptotic quantities such as the impulse --- see e.g.~Ref.~\cite{Jakobsen:2021zvh}.
The soft (S) region $\ell^\mu \sim (q\cdot v_1,q\cdot v_1)$ is more involved. We refer to the soft region of an integral by adding ``S" as a subscript of the respective integral.
In this region it suffices to obtain
$J^{1-\mathrm{loop}}_{0,1,0}$ and $J^{1-\mathrm{loop}}_{\mathrm{S},0,1,1}$.
The other two masters~\eqref{masterInts} are given by:
\begin{equation}
 J^{1-\mathrm{loop}}_{\mathrm{S},1,1,1}=-\frac{1}{\mathbf{q}^2} J^{1-\mathrm{loop}}_{\mathrm{S},0,1,1}\ , \qquad J^{1-\mathrm{loop}}_{\mathrm{S},1,1,0}=-\frac{1}{\mathbf{q}^2} J^{1-\mathrm{loop}}_{\mathrm{S},0,1,0}\ .
\end{equation}
The master integral
$J^{1-\mathrm{loop}}_{0,1,0}$ is a tadpole integral, which is well known and
can easily be solved.
\begin{align}
J^{1-\mathrm{loop}}_{0,1,0}=\left(\frac{-1}{4\pi}\right)^{\frac{D-1}{2}}\Gamma\left(1-\frac{D-1}{2}\right)(-q\cdot v_1+i0)^{D-3}\, .
\end{align}
So we are left only with $J^{1-\mathrm{loop}}_{\mathrm{S},0,1,1}$ for which we need an expression.

To derive the boundary integral $J^{1-\mathrm{loop}}_{\mathrm{S},0,1,1}$
we Schwinger parametrize the worldline propagator.
The result is a one-dimensional parameter integral of a spherically symmetric Fourier transform:
\begin{equation}
   J^{1-\mathrm{loop}}_{\mathrm{S},0,1,1}=
   (q\cdot v_1-i0)^{D-4}\int_0^\infty\!\d s\,e^{-\i s}\int_\mathbf{l}\frac{e^{-\i s \mathbf{l}\cdot \mathbf{v}}}{\mathbf{l}^2-1}\, .
\end{equation}
The $D$-dimensional spherically symmetric Fourier integral
is related to a Hankel transform of order $D/2-1$:
\begin{equation}
s^{\frac{D}{2}-1}\mathcal{F}[f(r)](s)=
(2\pi)^{-\frac{D}{2}}\mathcal{H}_{\frac{D}{2}-1}[r^{\frac{D}{2}-1}f(r)](s)\, ,
\end{equation}
where $\mathcal{F}[f(r)](s)$ is the D-dimensional Fourier transform of $f(r)$.
$\mathcal{H}_{\nu}[F](s)$ is the Hankel transform of the function $F$ of order $\nu$ as a function of the variable $s$:
\begin{equation}
\mathcal{H}_{\nu}[F](s):=\int_0^\infty F(r)J_{\frac{D}{2}-1}(sr)r\d r\, ,
\end{equation}
with $J_\nu$ the order-$\nu$ Bessel function of the first kind.
The resulting one-parameter integral is then straightforwardly solved,
and the $\eps$-expanded result is given by:
\begin{equation}
J^{1-\mathrm{loop}}_{\mathrm{S},0,1,1}=\frac{\i |q|^{-2\epsilon}}{8\pi\gamma v}\left(-\frac{1}{\eps}-\i\pi-2\,\mathrm{arccosh}(\gamma)-2\log[ (\hat{q}\cdot v_1)^{-1}_N]\right)+\mathcal{O}(\eps)\ .
\end{equation}
The full $q\cdot v_1\to 0$ limit of the master integrals is given by the sum of both regions.

\subsubsection{Identification of boundary regions}\label{sec:regionIdent}

The two contributing regions~\eqref{mspaceRegions} to the integrals in Feynman parameter space
in the $q\cdot v_1\to 0$ limit were determined with the help of
the \emph{Mathematica} package {\tt asy.m} \cite{Pak:2010pt,Jantzen:2012mw}.
In the Feynman parameter representation $J^{1-\mathrm{loop}}_{1,1,1}$ is given by
\begin{equation}
\label{fpint}
J^{1-\mathrm{loop}}_{1,1,1}=(4\pi)^{-\frac{D-1}2} \Gamma\bigg(3-\frac{D-1}2\bigg)\int_0^\infty \d x_1\int_0^\infty \d x_2\int_0^\infty \d x_3\, \delta(1-x_1-x_2-x_3)\,\mathcal{U}^{4-D}\mathcal{F}^{\frac{D-1}2-3}\ ,
\end{equation}
where
\begin{equation}
\mathcal{U}=-(x_1+x_2),\qquad \mathcal{F}=\frac{(2 \mathbf{q} x_1-\mathbf{v} x_3)^2}{4}-(x_1+x_2)(\mathbf{q}^2 x_1-(q\cdot v_1)^2 x_2-(q\cdot v_1) x_3)\ .
\end{equation}
Without loss of generality, in this discussion, we drop the $\i0$ in the scaling of the regions. Using {\tt asy.m} \cite{Pak:2010pt,Jantzen:2012mw} we identify the two regions $(0,0,0)$ and $(2,0,1)$,
where the numbers characterize the scalings in $q\cdot v_1$ of each parameter $x_i$:
\begin{subequations}\label{paramRegions}
\begin{align}
   &(0,0,0): & &(x_1,x_2,x_3)\sim(1,1,1)\,,\\
   &(2,0,1): & &(x_1,x_2,x_3)\sim((q\cdot v_1)^2,1,q\cdot v_1)\,.
\end{align}
\end{subequations}
% These triples specify the scaling with $q\cdot v_1$ of the three Feynman parameters for the integral $J^{1-\mathrm{loop}}_{1,1,1}$ in the Feynman parameter representation.
% Hence we assume the parameters to scale as $x_1\sim (q\cdot v_1)^2$, $x_2\sim (q\cdot v_1)^0$, $x_3\sim (q\cdot v_1)^1$ in the first region and as $x_1\sim (q\cdot v_1)^0$, $x_2\sim (q\cdot v_1)^0$, $x_3\sim (q\cdot v_1)^0$ in the second region.\gustav{do we need to comment on this?}
In these two regions the $\mathcal{U}$ and $\mathcal{F}$ polynomials become
\begin{subequations}
\begin{align}
   &(0,0,0): & \mathcal{U}&\to-(x_1+x_2), & \mathcal{F}&\to-\mathbf{q}^2 x_1 x_2+\frac{\mathbf{v}^2 x_3^2}{4}\,,\\
   &(2,0,1): & \mathcal{U}&\to-x_2, & \mathcal{F}&\to(q\cdot v_1)^2 \left(-\mathbf{q}^2 x_1 x_2+x_2^2+x_2 x_3+\frac{\mathbf{v}^2 x_3^2}{4}\right)\ .
\end{align}
\end{subequations}
In the first case the delta function is unchanged,
and in the second $\delta(1-x_1-x_2-x_3)\to\delta(1-x_2)$.
The $q\cdot v_1\to 0$ limit of the complete integral is given by the sum of both regions:
\begin{equation}
\label{sumofregions}
\begin{split}
\lim_{q\cdot v_1\to 0} J^{1-\mathrm{loop}}_{1,1,1}&=(4 \pi)^{-\frac{(D-1)}{2}}  \Gamma\bigg(3-\frac{(D-1)}{2}\bigg)\int_0^\infty \d x_1\int_0^\infty \d x_2\int_0^\infty \d x_3\\
&\times \bigg[(q\cdot v_1)^{D-4}\delta(1-x_2)\,(-x_2)^{4-D} \left(-\mathbf{q}^2 x_1 x_2+x_2^2+x_2 x_3+\frac{\mathbf{v}^2 x_3^2}{4}\right)^{(D-1)/2-3}\\
&+\delta(1-x_1-x_2-x_3)(-x_1-x_2)^{4-D} \left (-\mathbf{q}^2 x_1 x_2+\frac{\mathbf{v}^2 x_3^2}{4}\right)^{(D-1)/2-3}\bigg]\ .
\end{split}
\end{equation}
This allows us to directly identify the momentum space regions \eqref{mspaceRegions} with the parameter space
regions~\eqref{paramRegions}.
Starting from the integrands \eqref{J111reg1}, \eqref{J111regzf},
applying Feynman parametrization yields
\begin{subequations}
\begin{align}
   J^{1-\mathrm{loop}}_{1,1,1}&\xrightarrow{q\cdot v_1\to0}\Gamma(3)\int \frac{\d^{D-1}\mathbf{l}}{(2\pi)^{D-1}}\int_0^\infty \d x_1\int_0^\infty \d x_2\int_0^\infty \d x_3\frac{\delta(1-x_1-x_2-x_3)}{\left[\mathbf{l}^2(x_1+x_2)+\mathbf{q}^2 x_1-\frac{(-2 \mathbf{q} x_1+\mathbf{v}x_3)^2}{4(x_1+x_2)}\right]^3}\,,\\
   J^{1-\mathrm{loop}}_{1,1,1}&\xrightarrow{q\cdot v_1\to0}(q\cdot v_1)^{D-4}\Gamma(3)\int \frac{\d^{D-1}\mathbf{l}}{(2\pi)^{D-1}}\int_0^\infty \d x_1\int_0^\infty \d x_2\int_0^\infty \d x_3\frac{\delta(1-x_1-x_2-x_3)}{\left[\mathbf{l}^2 x_2+\mathbf{q}^2 x_1-(x_2+x_3)-\frac{(\mathbf{v}x_3)^2}{4 x_2}\right]^3}\,,
\end{align}
\end{subequations}
in the same two regions respectively.
Evaluating the remaining $\mathbf{l}$ integrals yields the two components in Eq.~\eqref{sumofregions},
thus validating our claim.
Similar considerations apply for the other master integrals.

\end{widetext}

\bibliography{paper}
\bibliographystyle{jhep}

\end{document}